\documentclass[journal]{IEEEtran}

\usepackage{algorithm}
\usepackage{algpseudocode}
\usepackage{float}
\usepackage{cite} 
\usepackage{graphicx}
\usepackage{amsmath, amsfonts}
\usepackage{xfrac}
\usepackage{enumitem}
\usepackage{tabularx}
\usepackage{caption}
\usepackage[caption=false,font=footnotesize]{subfig}
\usepackage{dblfloatfix} 
\usepackage{kantlipsum} 

\usepackage[english]{babel}

\usepackage{amsthm}

\newcommand{\specialcell}[2][c]{%
 \begin{tabular}[#1]{@{}c@{}}#2\end{tabular}}

\usepackage {tikz}
\usetikzlibrary {positioning}
\usetikzlibrary{shapes.geometric}
\usetikzlibrary{backgrounds}
\usetikzlibrary{intersections}
\usetikzlibrary{calc}
\tikzset{vertex style/.style={
 draw=#1,
 thick,
 fill=#1!70,
 text=white,
 ellipse,
 minimum width=2cm,
 minimum height=0.75cm,
 font=\small,
 outer sep=3pt,
 },
}
\tikzstyle{startstop} = [rectangle, inner sep=0,rounded corners, minimum width=1.2cm, minimum height=.8cm,text centered, draw=black]
\tikzstyle{io} = [trapezium, inner sep=0,trapezium left angle=70, trapezium right angle=110, inner ysep=8pt, outer sep=0pt,
 minimum height=1.81mm, minimum width=0pt, minimum height=.8cm, text centered, draw=black]
\tikzstyle{process} = [rectangle, inner sep=0.1cm, minimum width=1cm, minimum height=.8cm, text centered, draw=black]
\tikzstyle{decision} = [diamond, aspect=2, text width=5em, inner sep=0pt, text centered, draw=black]
\tikzstyle{arrow} = [thick, draw, -latex] 

\begin{document}

\title{Running the Network Harder:\\Connection Provisioning under Resource Crunch}

\author{Rafael B. R. Louren\c{c}o\IEEEauthorrefmark{1}, Massimo Tornatore\IEEEauthorrefmark{1}\IEEEauthorrefmark{2}, Charles U. Martel\IEEEauthorrefmark{1}, and Biswanath Mukherjee\IEEEauthorrefmark{1}
\\\small\IEEEauthorrefmark{1}University of California, Davis, USA, E-mail:[rlourenco, cumartel, bmukherjee]@ucdavis.edu, \\\small\IEEEauthorrefmark{2}Politecnico di Milano, Italy, E-mail:massimo.tornator@polimi.it.}

\maketitle
\thispagestyle{plain}

\begin{abstract}
Traditionally, networks operate at a small fraction of their capacities; however, recent technologies, such as Software-Defined Networking, may let operators run their networks harder (i.e., at higher utilization levels). Higher utilization can increase the network operator's revenue, but this gain comes at a cost: daily traffic fluctuations and failures might occasionally overload the network. We call such situations \textit{Resource Crunch}. Dealing with Resource Crunch requires certain types of flexibility in the system. We focus on scenarios with flexible bandwidth requirements, e.g., some connections can tolerate lower bandwidth allocation. This may free capacity to provision new requests that would otherwise be blocked. For that, the network operator needs to make an informed decision, since reducing the bandwidth of a high-paying connection to allocate a low-value connection is not sensible. We propose a strategy to decide whether or not to provision a request (and which other connections to degrade) focusing on maximizing profits during Resource Crunch. To address this problem, we use an abstraction of the network state, called a Connection Adjacency Graph (CAG). We propose PROVISIONER, which integrates our CAG solution with an efficient Linear Program (LP). We compare our method to existing greedy approaches and to LP-only solutions, and show that our method outperforms them during Resource Crunch. 
\end{abstract}


\section{Introduction}

Communication networks are designed to provide high availability for existing connections and high acceptance rates for new service requests. Traditionally, this dual goal has been achieved by deploying significant excess capacity, commonly utilizing less than 40\% of the total capacity \cite{hong2013achieving, jain2013b4}. The importance of this surplus capacity is threefold: it provides redundancy for allocated connections, it accommodates traffic variations (both predictable daily fluctuations and unpredictable traffic surges, such as flash crowds), and it serves as a cushion for traffic growth. There is great interest in using excess capacity, since running the network harder (i.e., at higher utilization levels) creates opportunities to generate revenue from previously-idle assets \cite{Dikbiyik:2012:EEC:2182751.2182760}.

In this study, we define \textit{Resource Crunch} as situations in which the network is temporarily faced with a larger offered load than it can possibly carry. The events that can cause Resource Crunch are either sudden increases in offered traffic (i.e., request arrivals) and/or capacity reductions (i.e., due to failures). In traditional networks, if a Resource Crunch occurs, then some of the new service requests can't be provisioned. Thus, traditional networks rely on excess capacity to reduce the chances of Resource Crunch.

Recent technological advances, such as Software-Defined Networking (SDN), allow networks to be operated at higher levels of utilization, thanks to the new capabilities they provide. Some practical wide-area SDN networks today report average link utilization above 60\% \cite{hong2013achieving, jain2013b4, 7006776, Zhang:2017:GDI:3068707.3068746}. This higher utilization enables more revenue per unit of capacity when compared to traditional networks. As the forecast compound annual growth rate of Internet traffic is very high \cite{ciscofebruary}, many traditional networks will be pushed to higher levels of utilization to maintain profitability, which increases the chances of Resource Crunch occurring. 

\begin{figure}
 \centering
 \includegraphics[width=3.in]{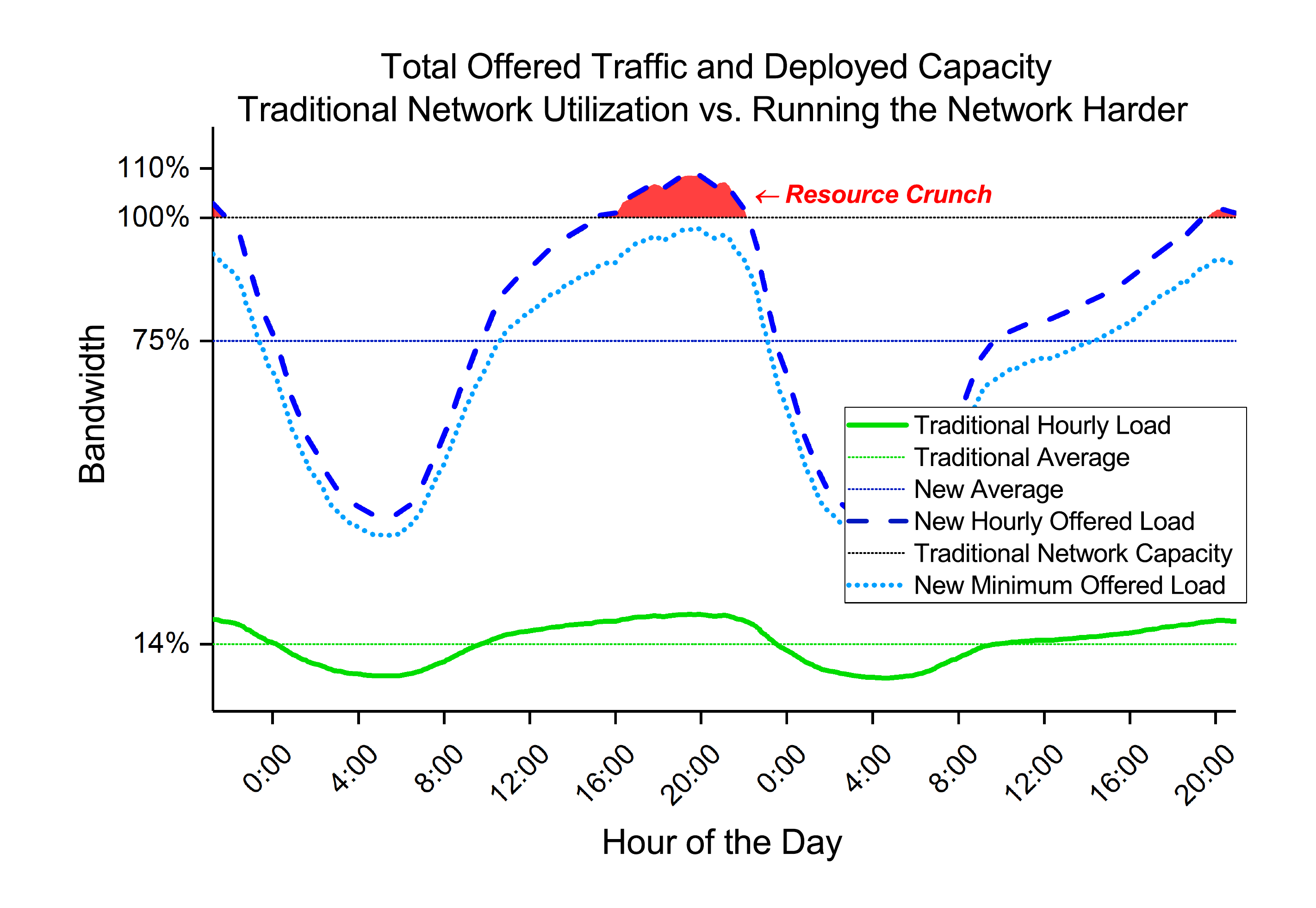}
 \caption{
 Aggregate daily traffic variation with respect to total network capacity. The green solid curve is based on real data collected from the Amsterdam Internet Exchange, from March 17 to 19, 2018. On these dates, the total capacity was of 27.46 Tbps while the average traffic carried on these dates was 3.73 Tbps (14\% of the total, and a peak-to-valley ratio of 2.91) \cite{amsix}. The upper blue dashed curve reproduces similar traffic variations, but at higher average utilization (75\%). The light-blue dotted curve represents the minimum requested traffic of such higher load.}	\label{fig:traditional-vs-new}
\end{figure}

An example detailing the fluctuations of offered/carried traffic through a period of two days is shown in Fig. \ref{fig:traditional-vs-new}. The green solid curve is the load carried by a traditional network. On average, this load remains at a low percentage of the total deployed capacity (in this case 14\%). The dark blue dashed curve shows what would happen if a similar fluctuation was true for a much larger load that increased the network utilization to 75\%. At this higher utilization, Resource Crunch (red shaded areas) can occur.

Resource Crunch can take up to a few hours to dissolve. This is an intrinsic difference between a network that is constantly congested and a network that goes through Resource Crunch from time to time. In the case of a congested network, the most appropriate solution would be to upgrade the network. However, a network that faces sporadic Resource Crunch can deal with them by exploring system flexibilities (e.g., malleable bandwidths, flexible start/end times, etc.) along with innovative routing strategies, as we will investigate. 

In this study, we consider Service Level Agreements (SLAs) with malleable bandwidth requirements. In that sense, requests ask for a certain bandwidth if it is available, but might be satisfied with some smaller bandwidth, down to a predefined minimum. We consider that this is true for the duration of the connection. For example, a connection might be initially allocated a certain bandwidth and, some time later, undergo a degradation which we call \textit{bandwidth throttling} to make room for an incoming request. Referring to the high network utilization example of Fig. \ref{fig:traditional-vs-new} (i.e., where the network is offered the load shown by the dark-blue dashed curve), the light-blue dotted curve illustrates the total minimum required bandwidth of the offered requests. Note that this curve never goes above the total deployed capacity. This indicates that, during Resource Crunch, there may be a way to serve all offered requests, at the expense of degrading some of them.

During a Resource Crunch, when a new request arrives and the network can't provision it, we call it a \textbf{\textit{crunched request}}. When a request is crunched, the network operator has to decide whether or not to serve this request. If the operator decides to serve it, the operator will need to degrade existing connections. If the operator blocks it, there might be penalties ranging from negative impacts on revenue (e.g., pre-agreed SLA penalties) to damages to the company's reputation.

In our study, we assume that each connection generates revenue according to some connection-specific function, which, among other things, depends on the connection's allocated bandwidth. Thus, when deciding whether or not to serve a crunched request, the following must be observed in order to maximize the operator’s profits during Resource Crunch (understood as the revenue generated from served connections, minus the penalties from blocking requests):
\begin{enumerate}
 \item Throttle connections only if the revenue loss due to those degradations is less than the revenue increase from the crunched request (plus the revenue saved from not having to pay its blocking penalty); 
 \item Throttle connections only if that will free enough bandwidth in order to allocate the crunched request; and 
 \item Select the cheapest possible set of connections to degrade, as throttling other connections would incur higher revenue losses. 
\end{enumerate}

With the observations above, we formulate a method to decide whether or not to serve a crunched request which also \textit{(i)} chooses which other connections to throttle, and \textit{(ii)} selects the path through which to route the crunched request. Our method uses a novel representation of the network state: \textit{Connection Adjacency Graph (CAG)}. This graph describes how allocated connections interact and how much revenue is lost when throttling any of them. We also propose an efficient Linear Program (LP) model to find a set of connections to degrade. We integrate these two tools into PROVISIONER, our proposed solution to solve this problem. Our illustrative results show that our method can achieve higher profits than an LP-only approach (and a heuristic) while providing service even for low-revenue requests.

The rest of this study is organized as follows: in Section \ref{sec:related}, we review related literature; in Section \ref{sec:rescrunch}, we propose a general representation of the revenue generated by bandwidth-flexible requests; in Section \ref{sec:problem}, we study the problem of connection provisioning with degradation during Resource Crunch, and introduce the CAG; in Section \ref{sec:cag-and-alg}, we introduce PROVISIONER; in Section \ref{sec:complexity}, we analyze the complexity of the CAG; we present illustrative results in Section \ref{sec:results}; and we conclude in Section \ref{sec:conclude}.

\section{Related Work} \label{sec:related}

The idea of running the network at high average link utilization has been studied through different perspectives. Achieving high utilization is a motivation to use certain technologies, such as SDN \cite{hong2013achieving}. This study, however, tackles the problem from an implementation perspective. Its approach is based on filling the excess capacity with low-priority traffic that can be dropped if necessary, hence, carrying the same amount of high-priority traffic as a traditional network would.

Some studies have analyzed the problem of provisioning requests to achieve some utility maximization. In \cite{Kumar:2015:BFH:2785956.2787478, jain2013b4} the goal is to maximize an abstract measure called \textit{fair share}. In \cite {7006776, Zhang:2017:GDI:3068707.3068746}, methods to fulfill inter-DC transfers prior to certain deadlines was studied. The problem of, given a set of requests, a network, and some objective (e.g., availability, revenue, etc.), how to serve the requests such that the objective is maximized was also studied by \cite{chiang2007layering, lin2006utility}. Our work differs from these studies in the following ways:
\begin{enumerate}
 \item Our method does not focus on reaching high utilization, per se. Instead, we look into what to do if Resource Crunch occurs;
 \item Our approach tries to provide connectivity as soon as a request arrives, without scheduling it for the future;
 \item Differently from \cite{Kumar:2015:BFH:2785956.2787478, jain2013b4}, we aim at having a joint decision of path and bandwidth allocation;
\item Requests arrive at any time and must be served as soon as possible, hence, we need a solution that works dynamically, instead of statically (as in \cite{chiang2007layering, lin2006utility});
 \item We do not re-route existing connections; and
 \item We consider requests with bandwidth flexibility.
\end{enumerate}

We assume that connections' can be throttled if necessary, i.e., they are malleable. Thus, our problem has similarities with the topic of \textit{malleable reservation} \cite{1250324}. Works on this topic generally focus on \textit{immediate reservation (IR)}, \textit{advance reservation (AR)} (also referred to as book-ahead reservation), or both. The book-ahead scheme was investigated by several works, such as \cite{665112,759312, 4351893}. Our study does not qualify as AR because our connections are either served or blocked as they come, without scheduling future allocations\footnote{In fact, we consider a continuous, non-slotted time dimension.}. However, book-ahead capability can be added as a future extension of our work. In \cite{Lu:15}, malleable reservation was studied in the context of Elastic Optical Networks, and a method to execute AR and IR for elastic optical channels was proposed.

In general, malleable reservation studies allow low priority and/or immediately reserved services be preempted (i.e., dropped) if a previously scheduled service needs the network resources that such a connection is utilizing. To the best of our knowledge, these works have not focused on throttling these connections instead of completely dropping them --- which is what we do in our study. Moreover, congestion is mostly considered in terms of how to avoid/minimize it instead of how to deal with unavoidable ones. We, however, focus on Resource Crunch, which causes an inevitable congestion that can last up to a few hours. That is a crucial difference because it allows us to use a non-shortest path based approach which can generate higher profits than shortest-path based approaches during Resource Crunch.

Similar problems to ours have been analyzed in other works, however, without our goal of maximizing the network operator's profits. In \cite{savas2016backup}, service degradation was used to reduce blocking and increase network survivability. Flexibility in both time-to-complete and desired bandwidth was studied in \cite{zhang2010reliable}, to allow for reliable multi-path provisioning, and in \cite{andrei2010provisioning} for deadline-driven scheduling. Considering the optical and electrical layers, \cite{gkamas2015joint} proposed a service degradation scheme using multi-path routing for minimum-cost network design, and \cite{7842043} analyzed the problem of provisioning degraded services while assuring Quality of Service (QoS). The algorithm proposed in \cite{7842043} is similar to \cite{roy2014network}; however, the latter aims at maximizing the network operator's revenue. These algorithms greedily search for a shortest path and serve a crunched request by executing degradations in that path from a lower priority (e.g., cheapest) to a higher priority (e.g., more expensive). Because our work aims at maximizing the network operator's profits, this approach is not always ideal, as will be further investigated in Section \ref{sec:results}. 

Our investigation also relates to admission control literature. In \cite{7744684}, the authors propose a dynamic-programming-based approximation to solve the dual problem of routing connections and provisioning their bandwidth, focusing on video-streaming services in an SDN environment. The authors provision network resources to incoming requests while maximizing revenues in the long run, given expected video-transmission rates. To the best of our knowledge, this is the only related work that also jointly focuses on admission control and routing of flows in wired networks. The solution presented does not deal with congestion scenarios such as Resource Crunch (when the offered load is larger than capacity). 


\section{Provisioning Requests: Profits and Resource Crunch}\label{sec:rescrunch}

\subsection{Profits}

Our study maximizes profits measured by the revenue from served connections minus the costs from blocked requests (i.e., we do not consider notions such as Capital and Operational Expenditures). We consider that, when a request is served, it generates revenue, but if it is blocked, it \textit{may} result in a penalty. This penalty can be due to many reasons (e.g., pre-agreed contract where the operator is expected to always provide connectivity for a certain client, etc.). 
For a request $d_i$, we define this blocking cost as:
$$F_i \in \mathbb{R}_{\geq 0} \,,$$
which is zero if such penalties do not exist.

To serve a request $d_i$, a new connection $c_i$ must be allocated.
The revenue generated by a connection $c_i$ may depend on several factors, such as:
\begin{itemize}
\item connection's bandwidth;
\item connection's service class (e.g., QoS or SLA); 
\item path-length/distance of $c_i$'s source and destination; 
\item popularity of route; 
\item time of the day (or day of the year); 
\item competition; 
\item regulation and taxation; among others.
\end{itemize}

In fact, each operator might use a different pricing strategy for different connectivity requests. For our study, we consider that we can describe the revenue generated by an allocated connection $c_i$ as a function of its bandwidth, defined by:
$$r_i:b \rightarrow r_i(b) \,,$$
for $b \in \mathbb{R}_{\geq 0}$. Thus, at any instant, we can use $r_i$ to compute how much revenue $c_i$ would generate for a different $b$.

We denote the possible bandwidths that connection $c_i$ might use as the interval $[B_i^{min}, B_i^{req}]$ (where $B_i^{min}$ is the minimum acceptable and $B_i^{req}$ is the requested bandwidth for that connection). This interval is particular to each request, and it may vary from $B_i^{min} \ll B_i^{req}$ ($c_i$ can be significantly throttled) to $B_i^{min} = B_i^{req}$ ($c_i$ cannot be throttled at all). 

As an example, in Section \ref{sec:results}, we use a pricing model that accounts for the length of the shortest path from the request's source to its destination, request's bandwidth, and its service class. The lowest priority classes of service use $B_i^{min} < B_i^{req}$; and the highest priority class uses $B_i^{min} = B_i^{req}$. 




\subsection{Resource Crunch}
In a network being run harder, a Resource Crunch can occur when a new request arrives or a failure occurs.

\begin{enumerate}
\item \textbf{Arrival of New Requests:} As a new request arrives, the total offered load to the network increases. Request arrivals can be categorized as:
\begin{enumerate}[label=\textit{\alph*.}]
\item\textit{Hourly/Daily Traffic Fluctuations:} On an hourly scale, the offered traffic changes according to time of the day, as shown in Fig. \ref{fig:traditional-vs-new} and explored by several studies \cite{ager2012anatomy, labovitz2010internet}. These changes have a periodic-like behavior, roughly repeating every day. Their peaks may generate a Resource Crunch (see red areas of Fig. \ref{fig:traditional-vs-new}) that lasts until the offered traffic decreases.

\item\textit{Traffic Surges:} This study is also applicable for sudden traffic peaks generated by unpredictable increases in traffic, such as those due to flash crowds.

\item\textit{Traffic Growth:} Resource Crunch can also occur when network upgrades don't keep up with traffic growth. Though our methods could help mitigate such a long-duration Resource Crunch, this is not the focus of this study and is an open problem for future research. 
\end{enumerate}
\item \textbf{Failures:} Failures can cause Resource Crunch even if the offered load remains the same since failures reduce the available capacity (e.g., lowering the black line of Fig. \ref{fig:traditional-vs-new}). Traditional networks are typically engineered to endure some (usually simple) failures. However, in a network running at high utilization, failures and disasters might cause a Resource Crunch. 
\end{enumerate}

When a request $d_c$ (with bandwidth interval $[B_c^{min}, B_c^{req}]$, and blocking penalty $F_c$) is crunched, we assume that, even before allocating it, we can describe its revenue function as:
$$r_{c}:b \rightarrow r_{c}(b) \,.$$ 
Depending on the pricing model, this might require approximations (for example, if the pricing model depends on which path the request is allocated). However, that is not always the case (as in the revenue model used in Section \ref{sec:results}).

\section{Problem Formulation}\label{sec:problem}

In this section, we state the problem of Provisioning under Resource Crunch (Section \ref{ssec:problem-statement}); provide some definitions that will be utilized in this study (Section \ref{ssec:definitions}); use an example to demonstrate different approaches to solve the problem (Section \ref{ssec:sp_approach}); and introduce the Connection Adjacency Graph.

\subsection{Problem Statement} \label{ssec:problem-statement}

\noindent\textit{Provisioning under Resource Crunch (PRC)}
\begin{enumerate}[label=\textit{\Roman*.}]
 \item \textit{Given:}
 \begin{itemize}
 \item Network topology: nodes, links, capacities;
 \item A set of allocated connections: revenue functions, paths, requested and minimum bandwidths; and
 \item A \textit{crunched request}: bandwidth, source-destination nodes, revenue function, blocking cost.
 \end{itemize}
 \item \textit{Output:} Whether or not to serve the request. If yes, then, also, a set of other connections to be throttled, how much to degrade each of them, and the path on which to place the crunched request.
 \item \textit{Constraints:} Link capacities, connections' minimum bandwidths, crunched request bandwidth interval, all connections are non-splittable.
\end{enumerate}

Our study considers a special version of the problem, focusing on \textbf{\textit{Profit Maximization Provisioning under Resource Crunch (PM-PRC)}}. Our goal is to \textit{maximize profits, measured by revenue generated from served connections after subtracting the cost of blocking requests.} We simplify the problem by trying to allocate crunched requests $d_c$ either at their minimum required bandwidth $B_{c}^{min}$ or at their requested bandwidth $B_{c}^{req}$, whichever generates more revenue (further explored in the next sections).

\subsection{Definitions} \label{ssec:definitions}

For a crunched request to be served, the following conditions must be met:
\begin{enumerate}[label=\textit{\alph*.}]
 \item A set of allocated connections must be found such that, after degrading these connections, there is a path from the crunched request's source to its destination with at least the minimum requested bandwidth of the crunched request (i.e., service degradations create a \textbf{\textit{freed path}} in the network).
 We call such set of allocated connections a \textbf{\textit{candidate degradation set}};
 \item Serving that request by executing these service degradations must be a \textbf{\textit{Profitable}} decision.
\end{enumerate}

We refer to the bandwidth that can be freed-up from a connection $c_i$ as its \textbf{\textit{degradable bandwidth}}. This is the difference between the current $B_{i}$ and $B_{i}^{min}$. Accordingly, for a connection $c_i$ we define the revenue lost by throttling its bandwidth from $B_{i}$ to $b$ as:
$$\widehat{r_{i}}(b) = \max (r_{i}(B_{i}) - r_{i}(b), 0) \,.$$

We define \textbf{\textit{Profitable}} as the positive result of the following:
\begin{enumerate}[label=\textit{\roman*.}]
 \item \textit{Given:}
 \begin{itemize}
 \item Crunched request $d_c$, its potential revenue increases $r_{c}(B_{c}^{req})$ and $r_{c}(B_{c}^{min})$, and its blocking cost $F_{c}$;
 \item Candidate degradation set $S^{min}$ of connections $c_i$ and the revenue that would be lost with it:
 $$L(S^{min}) = \sum\limits_{{c_i} \in S^{min}}\left(\widehat{r_{i}}(B_{i} - \delta^{min}_{i})\right)\,,$$ 
 where $\delta^{min}_{i}$ is the amount that $c_i$ will need to be degraded and $B_{i}$ is the currently allocated bandwidth of $c_i$. Note that, for any link in the path where $d_c$ will be allocated, the summation of all $\delta^{min}_{i}$ (and free capacity in the link) is at least equals to $B_{c}^{min}$. $L(S)$ is also called \textbf{\textit{degradation cost}}.
 \item Candidate degradation set $S^{req}$ of connections $c_i$ and the revenue that would be lost with it:
 $$L(S^{req}) = \sum\limits_{{c_i} \in S^{req}}\left(\widehat{r_{i}}(B_{i} - \delta^{req}_{i})\right)\,,$$ 
 where $\delta^{req}_{i}$ is the amount that $c_i$ will need to be degraded. Note that, for any link in the path where $d_c$ will be allocated, the summation of all $\delta^{req}_{i}$ (and free capacity in the link) is at least $B_{c}^{req}$.
 \end{itemize}
 \item \textit{Decide whether:} 
 $$\left[r_{c}(B_{c}^{min}) + F_{c}\right] \geq L(S^{min})\,,$$
 or
 $$\left[r_{c}(B_{c}^{req}) + F_{c}\right] \geq L(S^{req})\,.$$
 If both conditions above are true, check if
 \begin{align}
 \frac{L(S^{req})}{B_{c}^{req}} \leq \frac{L(S^{min})}{B_{c}^{min}}\,, \label{equation_revloss}
 \end{align}
 if so, choose to throttle $S^{req}$ and allocate the crunched request at $B_{c}^{req}$ bandwidth; otherwise, throttle $S^{min}$ and allocate the request at $B_{c}^{min}$.
\end{enumerate}

We define the \textbf{\textit{optimum candidate degradation set}} $S^{opt}$ as the candidate set that, among all possible candidate sets, for a given bandwidth (i.e., either $B_{c}^{min}$ or $B_{c}^{req}$), reduces the operator's revenue the least. $L(S^{opt})$ is the smallest across all possible degradation sets for a given bandwidth.

The \textit{optimum candidate degradation set} allows for the best \textit{Profitable} decision to be made. In other words, since we assume $r_{c}(B_{c}^{min})$, $r_{c}(B_{c}^{req})$, and $F_{c}$ of crunched request $d_c$ are predetermined, the optimum candidate degradation set enables the highest profits --- which is the goal of our study.



\subsection{Different Approaches to Solve the Problem}\label{ssec:sp_approach}

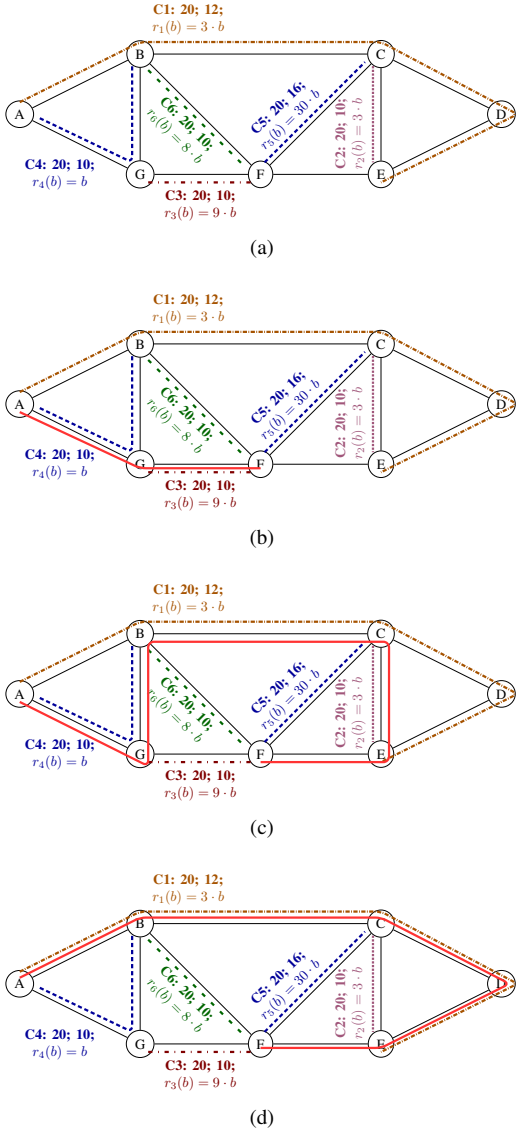
\begin{figure} [t]
\centering
\subfloat[\label{sfig:init}] {
\resizebox{!}{1.2in}{
 \begin {tikzpicture}[auto ,on grid , semithick,state/.style ={ ellipse, draw, text=black, align=center }]
 \node[state] (A) 													{A};
 \node[state] (B) 		[above right=1.5cm and 3cm of A] 			{B};
 \node[state] (C) 		[above right=1.5cm and 9cm of A]			{C};
 \node[state] (D) 		[right=12cm of A]							{D};
 \node[state] (E) 		[below right=1.5cm and 9cm of A]			{E};
 \node[state] (F) 		[below right=1.5cm and 6cm of A]			{F};
 \node[state] (G) 		[below right=1.5cm and 3cm of A]			{G};
 
 \node[color=black!40!blue, font=\bf] (C5new) 	[below right=1.5cm and 1cm of A]			{\specialcell[c]{C4: 20; 10;\\$r_4(b) = b$}};
 \node[color=black!40!blue, rotate=45, font=\bf] (C6new) 	[right=6.6cm of A]			{\specialcell[c]{C5: 20; 16;\\$r_5(b) = 30 \cdot b$}};
 \node[color=black!60!green, rotate=-45, font=\bf] (C7new) 	[below right=0.4cm and 4cm of A]			{\specialcell[c]{C6: 20; 10;\\$r_6(b) = 8 \cdot b$}};
 
 \path (A) 	edge 		node 	[above, sloped]	{}	(B)
				edge		node 	[below, sloped]	{}	(G)
	  (B) 	edge		node	[above, sloped]	{}	(C)
				edge		node	[above, sloped]	{}	(G)
	  (C) 	edge		node	[above, sloped]	{}	(D)
	 			edge		node	[above, sloped]	{}	(E)
	 	 (D) 	edge		node	[below, sloped]	{}	(E)
   (B) 	edge		node	[below, sloped]	{}	(F)
   (C) 	edge		node	[below, sloped]	{}	(F)
   (E) 	edge		node	[below, sloped]	{}	(F)
   (F) 	edge		node	[below, sloped]	{}	(G)
	 ;
	 
 \draw[color=black!35!orange, densely dashdotted, rounded corners,ultra thick] (0,0.3) -- (3,1.8) -- (9,1.8) node [above, pos = 0.2,font=\bf](C1) {\specialcell[c]{C1: 20; 12;\\$r_1(b) = 3\cdot b$}} -- (12.4, 0) -- (9.,-1.7) ;
 \draw[color=black!40!magenta, densely dotted, rounded corners,ultra thick] (8.8,1.2) -- (8.8,-1.2) node [above, rotate=90, pos = 0.7,font=\bf](C2) {\specialcell[c]{C2: 20; 10;\\$r_2(b) = 3\cdot b$}};
 \draw[color=black!50!red, loosely dashdotted, rounded corners,ultra thick] (3.2,-1.7) -- (5.8, -1.7) node [below, pos = 0.5,font=\bf](C4) {\specialcell[c]{C3: 20; 10;\\$r_3(b) = 9\cdot b$}};
 \draw[color=black!40!blue, densely dashed, rounded corners, ultra thick] (0.5,-0.1) -- (2.8, -1.2) -- (2.8,1.2);
 \draw[color=black!60!green, loosely dashed, rounded corners,ultra thick] (3.2, 1.1) -- (5.6, -1.3);
 \draw[color=black!40!blue, densely dashed, rounded corners, ultra thick] (6.1, -1.2) -- (8.6,1.3);
 \end{tikzpicture}
 }
} 

\subfloat[\label{sfig:shortpath}] {
 \resizebox{!}{1.2in}{
 \begin {tikzpicture}[auto ,on grid , semithick,state/.style ={ ellipse, draw, text=black, align=center }]
 \node[state] (A) 													{A};
 \node[state] (B) 		[above right=1.5cm and 3cm of A] 			{B};
 \node[state] (C) 		[above right=1.5cm and 9cm of A]			{C};
 \node[state] (D) 		[right=12cm of A]							{D};
 \node[state] (E) 		[below right=1.5cm and 9cm of A]			{E};
 \node[state] (F) 		[below right=1.5cm and 6cm of A]			{F};
 \node[state] (G) 		[below right=1.5cm and 3cm of A]			{G};
 
 \node[color=black!40!blue, font=\bf] (C5new) 	[below right=1.5cm and 1cm of A]			{\specialcell[c]{C4: 20; 10;\\$r_4(b) = b$}};
 \node[color=black!40!blue, rotate=45, font=\bf] (C6new) 	[right=6.6cm of A]			{\specialcell[c]{C5: 20; 16;\\$r_5(b) = 30 \cdot b$}};
 \node[color=black!60!green, rotate=-45, font=\bf] (C7new) 	[below right=0.4cm and 4cm of A]			{\specialcell[c]{C6: 20; 10;\\$r_6(b) = 8 \cdot b$}};
 
 \path (A) 	edge 		node 	[above, sloped]	{}	(B)
				edge		node 	[below, sloped]	{}	(G)
	  (B) 	edge		node	[above, sloped]	{}	(C)
				edge		node	[above, sloped]	{}	(G)
	  (C) 	edge		node	[above, sloped]	{}	(D)
	 			edge		node	[above, sloped]	{}	(E)
	 	 (D) 	edge		node	[below, sloped]	{}	(E)
   (B) 	edge		node	[below, sloped]	{}	(F)
   (C) 	edge		node	[below, sloped]	{}	(F)
   (E) 	edge		node	[below, sloped]	{}	(F)
   (F) 	edge		node	[below, sloped]	{}	(G)
	 ;
	 
 \draw[color=black!35!orange, densely dashdotted, rounded corners,ultra thick] (0,0.3) -- (3,1.8) -- (9,1.8) node [above, pos = 0.2,font=\bf](C1) {\specialcell[c]{C1: 20; 12;\\$r_1(b) = 3\cdot b$}} -- (12.4, 0) -- (9.,-1.7) ;
 \draw[color=black!40!magenta, densely dotted, rounded corners,ultra thick] (8.8,1.2) -- (8.8,-1.2) node [above, rotate=90, pos = 0.7,font=\bf](C2) {\specialcell[c]{C2: 20; 10;\\$r_2(b) = 3\cdot b$}};
 \draw[color=black!50!red, loosely dashdotted, rounded corners,ultra thick] (3.2,-1.7) -- (5.8, -1.7) node [below, pos = 0.5,font=\bf](C4) {\specialcell[c]{C3: 20; 10;\\$r_3(b) = 9\cdot b$}};
 \draw[color=black!40!blue, densely dashed, rounded corners, ultra thick] (0.5,-0.1) -- (2.8, -1.2) -- (2.8,1.2);
 \draw[color=black!60!green, loosely dashed, rounded corners,ultra thick] (3.2, 1.1) -- (5.6, -1.3);
 \draw[color=black!40!blue, densely dashed, rounded corners, ultra thick] (6.1, -1.2) -- (8.6,1.3);

 \draw[red!80, solid,rounded corners, ultra thick] (0,-0.2) -- (3, -1.6) -- (6,-1.6);
 \end{tikzpicture}
 }
} 

\subfloat[\label{sfig:shortpath-weight}] {
 \resizebox{!}{1.2in}{
 \begin {tikzpicture}[auto ,on grid , semithick,state/.style ={ ellipse, draw, text=black, align=center }]
 \node[state] (A) 													{A};
 \node[state] (B) 		[above right=1.5cm and 3cm of A] 			{B};
 \node[state] (C) 		[above right=1.5cm and 9cm of A]			{C};
 \node[state] (D) 		[right=12cm of A]							{D};
 \node[state] (E) 		[below right=1.5cm and 9cm of A]			{E};
 \node[state] (F) 		[below right=1.5cm and 6cm of A]			{F};
 \node[state] (G) 		[below right=1.5cm and 3cm of A]			{G};
 
 \node[color=black!40!blue, font=\bf] (C5new) 	[below right=1.5cm and 1cm of A]			{\specialcell[c]{C4: 20; 10;\\$r_4(b) = b$}};
 \node[color=black!40!blue, rotate=45, font=\bf] (C6new) 	[right=6.6cm of A]			{\specialcell[c]{C5: 20; 16;\\$r_5(b) = 30 \cdot b$}};
 \node[color=black!60!green, rotate=-45, font=\bf] (C7new) 	[below right=0.4cm and 4cm of A]			{\specialcell[c]{C6: 20; 10;\\$r_6(b) = 8 \cdot b$}};
 
 \path (A) 	edge 		node 	[above, sloped]	{}	(B)
				edge		node 	[below, sloped]	{}	(G)
	  (B) 	edge		node	[above, sloped]	{}	(C)
				edge		node	[above, sloped]	{}	(G)
	  (C) 	edge		node	[above, sloped]	{}	(D)
	 			edge		node	[above, sloped]	{}	(E)
	 	 (D) 	edge		node	[below, sloped]	{}	(E)
   (B) 	edge		node	[below, sloped]	{}	(F)
   (C) 	edge		node	[below, sloped]	{}	(F)
   (E) 	edge		node	[below, sloped]	{}	(F)
   (F) 	edge		node	[below, sloped]	{}	(G)
	 ;
	 
 \draw[color=black!35!orange, densely dashdotted, rounded corners,ultra thick] (0,0.3) -- (3,1.8) -- (9,1.8) node [above, pos = 0.2,font=\bf](C1) {\specialcell[c]{C1: 20; 12;\\$r_1(b) = 3\cdot b$}} -- (12.4, 0) -- (9.,-1.7) ;
 \draw[color=black!40!magenta, densely dotted, rounded corners,ultra thick] (8.8,1.2) -- (8.8,-1.2) node [above, rotate=90, pos = 0.7,font=\bf](C2) {\specialcell[c]{C2: 20; 10;\\$r_2(b) = 3\cdot b$}};
 \draw[color=black!50!red, loosely dashdotted, rounded corners,ultra thick] (3.2,-1.7) -- (5.8, -1.7) node [below, pos = 0.5,font=\bf](C4) {\specialcell[c]{C3: 20; 10;\\$r_3(b) = 9\cdot b$}};
 \draw[color=black!40!blue, densely dashed, rounded corners, ultra thick] (0.5,-0.1) -- (2.8, -1.2) -- (2.8,1.2);
 \draw[color=black!60!green, loosely dashed, rounded corners,ultra thick] (3.2, 1.1) -- (5.6, -1.3);
 \draw[color=black!40!blue, densely dashed, rounded corners, ultra thick] (6.1, -1.2) -- (8.6,1.3);

 \draw[red!80, solid, rounded corners, ultra thick] (0,-0.2) -- (3.2, -1.8) -- (3.2, 1.3) -- (9.2,1.3) -- (9.2,-1.7) -- (6,-1.7);
 \end{tikzpicture}
 }
} 

\subfloat[\label{sfig:longpath}] {
 \resizebox{!}{1.2in}{
 \begin {tikzpicture}[auto ,on grid , semithick,state/.style ={ ellipse, draw, text=black, align=center }]
 \node[state] (A) 													{A};
 \node[state] (B) 		[above right=1.5cm and 3cm of A] 			{B};
 \node[state] (C) 		[above right=1.5cm and 9cm of A]			{C};
 \node[state] (D) 		[right=12cm of A]							{D};
 \node[state] (E) 		[below right=1.5cm and 9cm of A]			{E};
 \node[state] (F) 		[below right=1.5cm and 6cm of A]			{F};
 \node[state] (G) 		[below right=1.5cm and 3cm of A]			{G};
 
 \node[color=black!40!blue, font=\bf] (C5new) 	[below right=1.5cm and 1cm of A]			{\specialcell[c]{C4: 20; 10;\\$r_4(b) = b$}};
 \node[color=black!40!blue, rotate=45, font=\bf] (C6new) 	[right=6.6cm of A]			{\specialcell[c]{C5: 20; 16;\\$r_5(b) = 30 \cdot b$}};
 \node[color=black!60!green, rotate=-45, font=\bf] (C7new) 	[below right=0.4cm and 4cm of A]			{\specialcell[c]{C6: 20; 10;\\$r_6(b) = 8 \cdot b$}};
 
 \path (A) 	edge 		node 	[above, sloped]	{}	(B)
				edge		node 	[below, sloped]	{}	(G)
	  (B) 	edge		node	[above, sloped]	{}	(C)
				edge		node	[above, sloped]	{}	(G)
	  (C) 	edge		node	[above, sloped]	{}	(D)
	 			edge		node	[above, sloped]	{}	(E)
	 	 (D) 	edge		node	[below, sloped]	{}	(E)
   (B) 	edge		node	[below, sloped]	{}	(F)
   (C) 	edge		node	[below, sloped]	{}	(F)
   (E) 	edge		node	[below, sloped]	{}	(F)
   (F) 	edge		node	[below, sloped]	{}	(G)
	 ;
	 
 \draw[color=black!35!orange, densely dashdotted, rounded corners,ultra thick] (0,0.3) -- (3,1.8) -- (9,1.8) node [above, pos = 0.2,font=\bf](C1) {\specialcell[c]{C1: 20; 12;\\$r_1(b) = 3\cdot b$}} -- (12.4, 0) -- (9.,-1.7) ;
 \draw[color=black!40!magenta, densely dotted, rounded corners,ultra thick] (8.8,1.2) -- (8.8,-1.2) node [above, rotate=90, pos = 0.7,font=\bf](C2) {\specialcell[c]{C2: 20; 10;\\$r_2(b) = 3\cdot b$}};
 \draw[color=black!50!red, loosely dashdotted, rounded corners,ultra thick] (3.2,-1.7) -- (5.8, -1.7) node [below, pos = 0.5,font=\bf](C4) {\specialcell[c]{C3: 20; 10;\\$r_3(b) = 9\cdot b$}};
 \draw[color=black!40!blue, densely dashed, rounded corners, ultra thick] (0.5,-0.1) -- (2.8, -1.2) -- (2.8,1.2);
 \draw[color=black!60!green, loosely dashed, rounded corners,ultra thick] (3.2, 1.1) -- (5.6, -1.3);
 \draw[color=black!40!blue, densely dashed, rounded corners, ultra thick] (6.1, -1.2) -- (8.6,1.3);

 \draw[red!80, ultra thick, rounded corners, solid] (0,0.15) -- (3, 1.65) -- (9,1.65) -- (12.2, 0) -- (9,-1.6) -- (6,-1.6);
 \end{tikzpicture}
 }
}

\caption{In this illustrative example, the initial state of the network can be seen in \ref{sfig:init}. Each link has 20 Gbps capacity. Six connections are allocated, each of their labels are organized as \textit{``Connection Number: Allocated Bandwidth (Gbps); Minimum Bandwidth (Gbps); Revenue Function"}. A new request $d_c$ from A to F arrives and it is crunched because all possible paths are congested. This request asks for $B_{c}^{req} = 10$ Gbps, $B_{c}^{min} = 5$ Gbps, has revenue function $r_{c}(b) = 6\cdot b$, and has a blocking fine $F_{c}$ of \$15. 
}
\label{fig:first_examples}
\end{figure}

When a request is crunched, a possible approach to decide whether or not to serve it (and how to allocate it) is to find the (k-) shortest path(s) between its source and destination. Then, select the cheapest set of connections in that path such that, once degraded, a freed path would be created. If it is profitable, degrade such connections and serve the request. This is similar to the approach of \cite{savas2016backup, roy2014network, 7842043}.

This approach chooses the service degradations that will be executed based on what connections traverse some specific path (namely, one of the k-shortest). Such an approach may not perform very well, as can be seen in Fig. \ref{sfig:shortpath}. Consider that a crunched request from A to F arrives. The shortest path is shown in solid red. Considering the minimum bandwidth $B_{c}^{min} = 5$ Gbps, this approach does not find a profitable solution, since degrading 5-Gbps connections C3 and C4 would decrease the total revenue \$50 and the crunched request would only increase the revenue \$30 (while its blocking cost is \$15). Hence, by using the shortest path, the crunched request is blocked although a solution exists.

Alternatively, a shortest-path algorithm could be executed taking as weights the degradation costs of connections. In this case, the solution found is also inaccurate. This inaccuracy leads to blocking of crunched requests needlessly; and, when crunched requests are not blocked, it does not necessarily find a good solution. For example, consider executing Dijkstra's algorithm taking as weights the degradation costs. The final path found by such an algorithm is shown in solid red in Fig. \ref{sfig:shortpath-weight}. For the minimum bandwidth $B_{c}^{min} = 5$ Gbps, the total degradation cost of that path is \$35. Since the crunched request is offering \$30 and has a blocking fine of \$15, it would be allocated. However, if the blocking cost of the crunched request was zero, it would be rejected.

Note that, to allocate the crunched request at its minimum bandwidth $B_{c}^{min} = 5$ Gbps, the \textit{optimum candidate degradation set} consists of connection C1. Degrading C1 by 5 Gbps would decrease revenue by \$15, which is less than the revenue increase offered by the crunched request (without even considering its blocking cost). This degradation allows for the crunched request to be served by the solid red path of Fig. \ref{sfig:longpath}. 

In Section \ref{sec:algorithm}, we will introduce our method to solve this problem. Our solution is based on two observations: \textit{i)} we can select a path and try to degrade connections along that path; \textit{ii)} we can select connections to degrade and, based on those connections, find a path. To take care of the first observation, we use tools such as k-shortest path routing and Linear Programming (LP). To solve the second observation in the next Section we introduce a new tool, called Connection Adjacency Graph.

\subsection{Connection Adjacency Graph (CAG)}\label{sec:cag}

When we throttle an allocated connection, capacity is freed along that entire connection's path, and not necessarily only on the links we are interested in. Referring to the example of Fig. \ref{sfig:shortpath}, if connections C3 and C4 are degraded to make room for a crunched request from A to F, capacity will be freed not only through links A-G and G-F, but also on link B-G. Hence, rather than degrade services to free capacity throughout some predefined path (similar to the first approach of Section \ref{ssec:sp_approach}), it might be better to focus on finding a good set of connections that together can free up enough capacity for the new request. There are many potential sets of connections to consider, so we have to find a good way to focus on the better solutions.

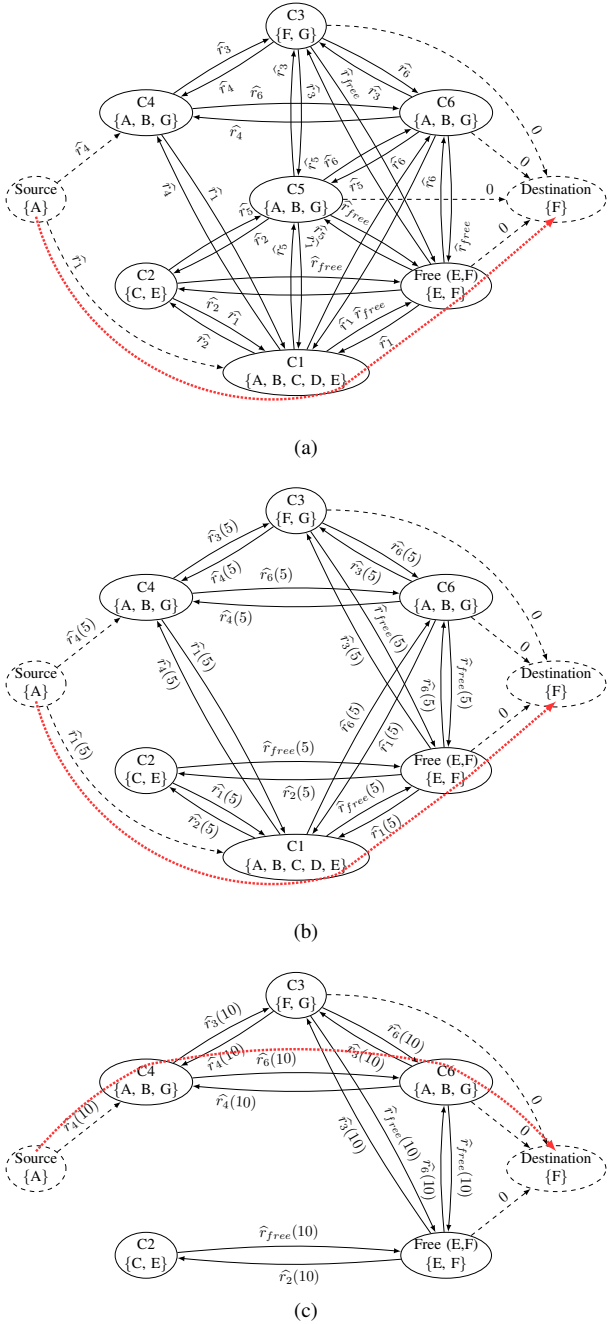
\begin{figure} [t]
\centering
\subfloat[\label{sfig:general_cag}] {
\resizebox{3.2in}{!}{
 \begin {tikzpicture}[-latex ,auto ,on grid , semithick,state/.style ={ ellipse, draw, text=black, align=center, inner sep=0pt}]
 	\def\VARSIZE{8}
 \def\HEIGHTDISCOUNT{1}
 \node[state] (C2) 	at (-1.73*\VARSIZE/4, -\VARSIZE*\HEIGHTDISCOUNT/4)		{C2\\ \{C, E\}};
 \node[state] (C1) 	at (0, -\VARSIZE*\HEIGHTDISCOUNT/2)						{C1\\ \{A, B, C, D, E\}};
 \node[state] (C3) 	at (0, \VARSIZE*\HEIGHTDISCOUNT/2)						{C3\\ \{F, G\}};
 \node[state] (C4) 	at (-1.73*\VARSIZE/4, \VARSIZE*\HEIGHTDISCOUNT/4)		{C4\\ \{A, B, G\}};
 \node[state] (C5) 	at (0,0)												{C5\\ \{A, B, G\}};
 \node[state] (C6) 	at (1.73*\VARSIZE/4, \VARSIZE*\HEIGHTDISCOUNT/4)		{C6\\ \{A, B, G\}};
 \node[state] (FL) 	at (1.73*\VARSIZE/4, -\VARSIZE*\HEIGHTDISCOUNT/4)		{Free (E,F)\\ \{E, F\}};
 \node[state, dashed] (S) 	at (-.75*\VARSIZE, 0)							{Source\\ \{A\}};
 \node[state, dashed] (T) 	at (.75*\VARSIZE, 0)							{Destination\\ \{F\}};
 
 \path (C1) 	edge 	[bend left=5]	node 	[below, sloped, pos=0.8]	{$\widehat{r_4}$}	(C4)
				edge	[bend left=5]	node 	[below, sloped, pos=0.5]	{$\widehat{r_2}$}	(C2)
    edge	[bend left=5]	node 	[above, sloped, pos=0.8]	{$\widehat{r_5}$}	(C5)
    edge	[bend left=5]	node 	[above, sloped, pos=0.85]	{$\widehat{r_6}$}	(C6)
    edge	[bend left=5]	node 	[above, sloped, pos=0.6]	{$\widehat{r}_{free}$}	(FL)
	 	(C2) 	edge	[bend left=5]	node	[above, sloped, pos=0.6]	{$\widehat{r_1}$}	(C1)
    edge	[bend left=5]	node	[above, sloped, pos=0.9]	{$\widehat{r_5}$}	(C5)
    edge	[bend left=5]	node	[above, sloped, pos=0.68]	{$\widehat{r}_{free}$}				(FL)
	 	(C3) 	edge	[bend left=5]	node	[below, sloped, pos=0.6]	{$\widehat{r_4}$}	(C4)
  		edge	[bend left=5]	node	[below, sloped, pos=0.9]	{$\widehat{r_5}$}	(C5)
  		edge	[bend left=5]	node	[above, sloped, pos=0.8]	{$\widehat{r_6}$}	(C6)
  		edge	[bend left=5]	node	[above, sloped, pos=0.2]	{$\widehat{r}_{free}$}				(FL)
    edge	[dashed, bend left=35]	node	[above, sloped, pos=0.85]	{$0$}	(T)
	 	(C4) 	edge	[bend left=5]	node	[above, sloped, pos=0.3]	{$\widehat{r_1}$}	(C1)
				edge	[bend left=5]	node	[above, sloped, pos=0.6]	{$\widehat{r_3}$}	(C3)
    edge	[bend left=5]	node	[above, sloped, pos=0.3]	{$\widehat{r_6}$}	(C6)
  	(C5) 	edge	[bend left=5]	node	[above, sloped, pos=0.15]	{$\widehat{r_1}$}	(C1)
				edge	[bend left=5]	node	[below, sloped, pos=0.15]	{$\widehat{r_2}$}	(C2)
    edge	[bend left=5]	node	[above, sloped, pos=0.85]	{$\widehat{r_3}$}	(C3)
    edge	[bend left=5]	node	[above, sloped, pos=0.15]	{$\widehat{r_6}$}	(C6)
    edge	[bend left=5]	node	[above, sloped, pos=0.2]	{$\widehat{r}_{free}$}				(FL)
    edge	[dashed]		node	[above, sloped, pos=0.9]	{$0$}	(T)
  	(C6) 	edge	[bend left=5]	node	[below, sloped, pos=0.85]	{$\widehat{r_1}$}	(C1)
				edge	[bend left=5]	node	[below, sloped, pos=0.3]	{$\widehat{r_3}$}	(C3)
    edge	[bend left=5]	node	[below, sloped, pos=0.8]	{$\widehat{r_4}$}	(C4)
    edge	[bend left=5]	node	[below, sloped, pos=0.8]	{$\widehat{r_5}$}	(C5)
    edge	[bend left=5]	node	[below, sloped, pos=0.8]	{$\widehat{r}_{free}$}				(FL)
    edge	[dashed]		node	[above, sloped, pos=0.8]	{$0$}	(T)
  (FL)	edge	[bend left=5]	node 	[below, sloped, pos=0.5]	{$\widehat{r_1}$}	(C1)
  		edge	[bend left=5]	node 	[below, sloped, pos=0.85]	{$\widehat{r_2}$}	(C2)
    edge	[bend left=5]	node	[below, sloped, pos=0.85]	{$\widehat{r_3}$}	(C3)
    edge	[bend left=5]	node	[below, sloped, pos=0.95]	{$\widehat{r_5}$}	(C5)
    edge	[bend left=5]	node	[above, sloped, pos=0.65]	{$\widehat{r_6}$}	(C6)
    edge	[dashed]		node	[above, sloped, pos=0.65]	{$0$}	(T)
  (S) 	edge	[dashed, bend right=30]	node	[above, sloped, pos=0.2]	{$\widehat{r_1}$}	(C1)
    edge	[dashed]				node	[above, sloped, pos=0.5]	{$\widehat{r_4}$}	(C4)
	 	;	
  
  \draw[red!80, ultra thick, rounded corners, densely dotted] (-.75*\VARSIZE, -.05*\VARSIZE) to [bend right=45] (.12*\VARSIZE, -1.1*\VARSIZE*\HEIGHTDISCOUNT/2) to (1.95*\VARSIZE/4, -1.1*\VARSIZE*\HEIGHTDISCOUNT/4) -- (.75*\VARSIZE, -.05*\VARSIZE);
 \end{tikzpicture}
 }
}

\subfloat[\label{sfig:min_cag}] {
 \resizebox{3.2in}{!}{
 \begin {tikzpicture}[-latex ,auto ,on grid , semithick,state/.style ={ ellipse, draw, text=black, align=center, inner sep=0pt}]
 	\def\VARSIZE{8}
 \def\HEIGHTDISCOUNT{1}
 \node[state] (C2) 	at (-1.73*\VARSIZE/4, -\VARSIZE*\HEIGHTDISCOUNT/4)		{C2\\ \{C, E\}};
 \node[state] (C1) 	at (0, -\VARSIZE*\HEIGHTDISCOUNT/2)						{C1\\ \{A, B, C, D, E\}};
 \node[state] (C3) 	at (0, \VARSIZE*\HEIGHTDISCOUNT/2)						{C3\\ \{F, G\}};
 \node[state] (C4) 	at (-1.73*\VARSIZE/4, \VARSIZE*\HEIGHTDISCOUNT/4)		{C4\\ \{A, B, G\}};
 \node[state] (C6) 	at (1.73*\VARSIZE/4, \VARSIZE*\HEIGHTDISCOUNT/4)		{C6\\ \{A, B, G\}};
 \node[state] (FL) 	at (1.73*\VARSIZE/4, -\VARSIZE*\HEIGHTDISCOUNT/4)		{Free (E,F)\\ \{E, F\}};
 \node[state, dashed] (S) 	at (-.75*\VARSIZE, 0)							{Source\\ \{A\}};
 \node[state, dashed] (T) 	at (.75*\VARSIZE, 0)							{Destination\\ \{F\}};
 
 \path (C1) 	edge 	[bend left=5]	node 	[below, sloped, pos=0.8]	{$\widehat{r_4}(5)$}	(C4)
				edge	[bend left=5]	node 	[below, sloped, pos=0.5]	{$\widehat{r_2}(5)$}	(C2)
    edge	[bend left=5]	node 	[above, sloped, pos=0.5]	{$\widehat{r_6}(5)$}	(C6)
    edge	[bend left=5]	node 	[above, sloped, pos=0.5]	{$\widehat{r}_{free}(5)$}	(FL)
	 	(C2) 	edge	[bend left=5]	node	[above, sloped, pos=0.5]	{$\widehat{r_1}(5)$}	(C1)
    edge	[bend left=5]	node	[above, sloped, pos=0.5]	{$\widehat{r}_{free}(5)$}				(FL)
	 	(C3) 	edge	[bend left=5]	node	[below, sloped, pos=0.6]	{$\widehat{r_4}(5)$}	(C4)
  		edge	[bend left=5]	node	[above, sloped, pos=0.8]	{$\widehat{r_6}(5)$}	(C6)
  		edge	[bend left=5]	node	[above, sloped, pos=0.5]	{$\widehat{r}_{free}(5)$}				(FL)
    edge	[dashed, bend left=35]	node	[above, sloped, pos=0.85]	{$0$}	(T)
	 	(C4) 	edge	[bend left=5]	node	[above, sloped, pos=0.2]	{$\widehat{r_1}(5)$}	(C1)
				edge	[bend left=5]	node	[above, sloped, pos=0.6]	{$\widehat{r_3}(5)$}	(C3)
    edge	[bend left=5]	node	[above, sloped, pos=0.4]	{$\widehat{r_6}(5)$}	(C6)
  	(C6) 	edge	[bend left=5]	node	[below, sloped, pos=0.5]	{$\widehat{r_1}(5)$}	(C1)
				edge	[bend left=5]	node	[below, sloped, pos=0.4]	{$\widehat{r_3}(5)$}	(C3)
    edge	[bend left=5]	node	[below, sloped, pos=0.8]	{$\widehat{r_4}(5)$}	(C4)
    edge	[bend left=5]	node	[above, sloped, pos=0.5]	{$\widehat{r}_{free}(5)$}				(FL)
    edge	[dashed]		node	[above, sloped, pos=0.8]	{$0$}	(T)
  (FL)	edge	[bend left=5]	node 	[below, sloped, pos=0.5]	{$\widehat{r_1}(5)$}	(C1)
  		edge	[bend left=5]	node 	[below, sloped, pos=0.45]	{$\widehat{r_2}(5)$}	(C2)
    edge	[bend left=5]	node	[below, sloped, pos=0.5]	{$\widehat{r_3}(5)$}	(C3)
    edge	[bend left=5]	node	[below, sloped, pos=0.4]	{$\widehat{r_6}(5)$}	(C6)
    edge	[dashed]		node	[above, sloped, pos=0.65]	{$0$}	(T)
  (S) 	edge	[dashed, bend right=30]	node	[above, sloped, pos=0.2]	{$\widehat{r_1}(5)$}	(C1)
    edge	[dashed]				node	[above, sloped, pos=0.5]	{$\widehat{r_4}(5)$}	(C4)
	 	;	

  \draw[red!80, ultra thick, rounded corners, densely dotted] (-.75*\VARSIZE, -.05*\VARSIZE) to [bend right=45] (.12*\VARSIZE, -1.1*\VARSIZE*\HEIGHTDISCOUNT/2) to (1.95*\VARSIZE/4, -1.1*\VARSIZE*\HEIGHTDISCOUNT/4) -- (.75*\VARSIZE, -.05*\VARSIZE);
 \end{tikzpicture}
 }
}

\subfloat[\label{sfig:req_cag}] {
 \resizebox{3.2in}{!}{
 \begin {tikzpicture}[-latex ,auto ,on grid , semithick,state/.style ={ ellipse, draw, text=black, align=center, inner sep=0pt}]
 	\def\VARSIZE{8}
 \def\HEIGHTDISCOUNT{1}
 \node[state] (C2) 	at (-1.73*\VARSIZE/4, -\VARSIZE*\HEIGHTDISCOUNT/4)		{C2\\ \{C, E\}};
 \node[state] (C3) 	at (0, \VARSIZE*\HEIGHTDISCOUNT/2)						{C3\\ \{F, G\}};
 \node[state] (C4) 	at (-1.73*\VARSIZE/4, \VARSIZE*\HEIGHTDISCOUNT/4)		{C4\\ \{A, B, G\}};
 \node[state] (C6) 	at (1.73*\VARSIZE/4, \VARSIZE*\HEIGHTDISCOUNT/4)		{C6\\ \{A, B, G\}};
 \node[state] (FL) 	at (1.73*\VARSIZE/4, -\VARSIZE*\HEIGHTDISCOUNT/4)		{Free (E,F)\\ \{E, F\}};
 \node[state, dashed] (S) 	at (-.75*\VARSIZE, 0)							{Source\\ \{A\}};
 \node[state, dashed] (T) 	at (.75*\VARSIZE, 0)							{Destination\\ \{F\}};
 
 \path 
  (C2) 	edge	[bend left=5]	node	[above, sloped, pos=0.5]	{$\widehat{r}_{free}(10)$}				(FL)
	 	(C3) 	edge	[bend left=5]	node	[below, sloped, pos=0.6]	{$\widehat{r_4}(10)$}	(C4)
  		edge	[bend left=5]	node	[above, sloped, pos=0.8]	{$\widehat{r_6}(10)$}	(C6)
  		edge	[bend left=5]	node	[above, sloped, pos=0.6]	{$\widehat{r}_{free}(10)$}				(FL)
    edge	[dashed, bend left=35]	node	[above, sloped, pos=0.85]	{$0$}	(T)
		(C4) 	edge	[bend left=5]	node	[above, sloped, pos=0.6]	{$\widehat{r_3}(10)$}	(C3)
    edge	[bend left=5]	node	[above, sloped, pos=0.4]	{$\widehat{r_6}(10)$}	(C6)
		(C6)	edge	[bend left=5]	node	[below, sloped, pos=0.4]	{$\widehat{r_3}(10)$}	(C3)
    edge	[bend left=5]	node	[below, sloped, pos=0.8]	{$\widehat{r_4}(10)$}	(C4)
    edge	[bend left=5]	node	[above, sloped, pos=0.5]	{$\widehat{r}_{free}(10)$}				(FL)
    edge	[dashed]		node	[above, sloped, pos=0.8]	{$0$}	(T)
  (FL)	edge	[bend left=5]	node 	[below, sloped, pos=0.45]	{$\widehat{r_2}(10)$}	(C2)
    edge	[bend left=5]	node	[below, sloped, pos=0.5]	{$\widehat{r_3}(10)$}	(C3)
    edge	[bend left=5]	node	[below, sloped, pos=0.4]	{$\widehat{r_6}(10)$}	(C6)
    edge	[dashed]		node	[above, sloped, pos=0.65]	{$0$}	(T)
  (S) 	
    edge	[dashed]				node	[above, sloped, pos=0.5]	{$\widehat{r_4}(10)$}	(C4)
	 	;	
  
  \draw[red!80, ultra thick, rounded corners, densely dotted] (-.75*\VARSIZE, .05*\VARSIZE) to [bend left = 10] (-1.73*\VARSIZE/4, 1.2*\VARSIZE*\HEIGHTDISCOUNT/4) to [bend left = 10] (1.73*\VARSIZE/4, 1.2*\VARSIZE*\HEIGHTDISCOUNT/4) to [bend left = 10] (.75*\VARSIZE, .05*\VARSIZE);
 \end{tikzpicture}
 }
}

\caption{\ref{sfig:general_cag} shows the relaxed $CAG(N, s, t)$ for network state $N$ of Fig. \ref{sfig:init}. \ref{sfig:min_cag} shows $CAG^{min}(N, s, t, B^{min}_{c})$. \ref{sfig:req_cag} shows $CAG^{req}(N, s, t, B^{req}_{c})$. Source and destination vertices are dashed (as well as their incoming/outgoing edges). Min-cost paths are shown by the dotted red paths.}
\label{fig:cag_examples}
\end{figure}

The Connection Adjacency Graph (CAG) is a directed, weighted graph that provides an abstract view of how connections interact, allowing for cheap degradation candidates to be found. In a network at a certain state $N$, for a crunched request with source $s$ and destination $t$, we generate $CAG(N, s, t)$ (also called the relaxed CAG) with the following steps:
\begin{enumerate}
 \item For each connection that has some degradable capacity, add a vertex to the CAG. Associate with that vertex a list of the physical nodes on that connection's path;
 \item For any link $(i,j)$ in the network which has free capacity, we create a dummy connection between $i$ and $j$ which has associated function $\widehat{r}_{free}(b)$, 
and add its respective vertex to the CAG. The functions $\widehat{r}_{free}(b)$ will provide useful flexibility in the next section. However, for now, consider that all $\widehat{r}_{free}(b) = 0$;
 \item For each pair of vertices $l,k$ in the CAG, add directed edges $(l,k)$ and $(k,l)$ if the connections associated with $l$ and $k$ have at least one physical node in common;
 \item Each incoming edge of $l$ will get its weight using the lost revenue function $\widehat{r_i}$ of connection $c_i$ associated with $l$;
 \item Create a dummy source vertex and add edges from it to each other vertex of a connection that includes node $s$ (as usual the edge weight is given using the function $\widehat{r_i}$ of the destination vertex). Similarly, create a dummy destination vertex, and add edges from each vertex of a connection containing $t$ to this dummy vertex (edges with weight $\widehat{r_{t}}(b)=0$).
\end{enumerate}

A path from $s$ to $t$ in the CAG represents a set of connections such that, if each of them were reduced by $B^{min}_c$, we would create a free path. Not all connections can be reduced by that much, but we can consider two \textit{relaxed} versions of the problem where any connection (including dummy ones) in the CAG can be degraded by $B^{min}_c$ (or by $B^{req}_c$). For that, the edge weights of the relaxed $CAG(N, s, t)$ are set using the lost revenue functions of the vertices that each edge points to. Thus, we define $CAG(N, s, t, B^{min}_{c})$ and $CAG(N, s, t, B^{req}_{c})$ such that the weights of each edge $(i,j)$ in each of them is given by $\widehat{r_j}(B^{min}_{c})$ and $\widehat{r_j}(B^{req}_{c})$, respectively. The weight of the edge represents the lost revenue if we reduced the connection that such edge points to by that amount.

$CAG(N, s, t)$ can be kept in memory and only updated as connections arrive/depart (i.e., as network state $N$ changes). When a connection is provisioned, it is necessary to add a vertex representing it in the CAG. It is also necessary to check if that connection exhausted any previously-free capacity on any of its links. In that case, the vertices that represented those free capacities must be removed from the CAG (after adding the new vertex). When a connection departs, it is necessary to check if it is freeing up capacity on any link that was previously fully occupied. In this case, it is necessary to add vertices representing the new free capacities to the CAG (while removing the vertex of the departing connection). 

The amount by which each connection may be throttled has not been accounted for so far --- i.e., this constraint has been relaxed. To account for it, we use two other graphs: $CAG^{min}(N, s, t, B^{min}_{c})$ and $CAG^{req}(N, s, t, B^{req}_{c})$. To generate each of them, for each vertex $i$ in $CAG(N, s, t)$: 
\begin{enumerate}
\item Find the minimum free capacity $f_{min}$ among all links of the connection that $i$ represents;
\item Compute the sum $v_{bw}$ of $f_{min}$ plus the degradable capacity of the connection associated to $i$;
\item If $v_{bw}$ is equals to or larger than $B^{min}_{c}$, copy $i$ and all of its incoming edges from $CAG(N, s, t, B^{min}_{c})$ to $CAG^{min}(N, s, t, B^{min}_{c})$;
\item If $v_{bw}$ is equals to or larger than $B^{req}_{c}$, copy $i$ and all of its incoming edges from $CAG(N, s, t, B^{req}_{c})$ to $CAG^{req}(N, s, t, B^{req}_{c})$.
\end{enumerate}


Illustrative examples of these graphs are shown in Fig. \ref{fig:cag_examples}, based on network state $N$ of Fig. \ref{sfig:init}. Notice how the physical network topology is abstracted in the CAG. The red dotted paths shown in Fig. \ref{fig:cag_examples} are:
\begin{itemize}
\item \textit{Fig. \ref{sfig:general_cag}}: min-cost $s-t$ path for $B^{min}_{c} = 5$ Gbps and $B^{req}_{c} = 10$ Gbps in $CAG(N, s, t, B^{min}_{c})$ and $CAG(N, s, t, B^{req}_{c})$, respectively (both paths coincide);
\item \textit{Fig. \ref{sfig:min_cag}}: min-cost $s-t$ path in $CAG^{min}(N, s, t, B^{min}_{c})$;
\item \textit{Fig. \ref{sfig:req_cag}}: min-cost $s-t$ path in $CAG^{req}(N, s, t, B^{req}_{c})$.
\end{itemize}

The vertices of the min-cost paths shown in Figs. \ref{sfig:min_cag} and \ref{sfig:req_cag} (except for Source and Destination) are the optimum degradation sets for $B^{min}_{c}$ and $B^{req}_{c}$, respectively. That is because in the example used for Figs. \ref{fig:first_examples} and \ref{fig:cag_examples} each link has at most one connection. Accordingly, our main interest in using the CAG is to find min-cost paths from Source to Destination\footnote{It is not necessary to effectively maintain different graphs for the relaxed and non-relaxed CAGs in memory. This is because the min-cost path algorithm can use just the relaxed $CAG(N, s, t)$ if, for $CAG^{min}(N, s, t, B^{min}_{c})$ and $CAG^{req}(N, s, t, B^{req}_{c})$, it ignores vertices and edges that should not be contained in each of these graphs. Also, as it executes, this algorithm can compute edge $(i,j)$ weight with $\widehat{r_j}(B^{min}_{c})$ or $\widehat{r_j}(B^{req}_{c})$.}. The min-cost path in Fig. \ref{sfig:min_cag} costs the same as that of Fig. \ref{sfig:general_cag} (for $B^{min}_{c} = 5$ Gbps). However, this is not true when comparing the path of Figs. \ref{sfig:req_cag} and \ref{sfig:general_cag} (for $B^{req}_{c} = 10$ Gbps). 

The costs of min-cost paths found in $CAG(N, s, t, B^{min}_{c})$ and $CAG(N, s, t, B^{req}_{c})$ are lower bounds on the costs of min-cost paths in $CAG^{min}(N, s, t, B^{min}_{c})$ and $CAG^{req}(N, s, t, B^{req}_{c})$, respectively. This is because these graphs are contained in the relaxed CAGs; thus, all paths that can be found in them can also be found in relaxed CAGs. However, a path found in the relaxed CAG might not have enough capacity to serve a certain request (as the degradable capacity constraint is not considered in the relaxed CAGs) --- which is not true for non-relaxed CAGs. If the min-cost path in either $CAG^{min}(N, s, t, B^{min}_{c})$ or $CAG^{req}(N, s, t, B^{req}_{c})$ costs the same as the min-cost path in the respective relaxed CAG we call it a \textit{good candidate set} $S^{good}$. 

With the information provided by the min-cost path found through the CAG, we can proceed to degrade the connections represented by the vertices of that path if that is Profitable. When these connections are degraded, capacity will be freed in the network from the crunched request's source to its destination. Hence, we need to perform another shortest-path computation, but this time, on the actual network to find the path through which the request will be allocated. Note that such path might have different amounts of free capacity in each link; thus, after allocating the crunched request, we might be able to re-upgrade some of the connections that were just throttled (reducing even more the lost revenue).

In more realistic scenarios than that of Fig. \ref{fig:first_examples}, links might be traversed by multiple connections. Thus, it is not always true that the min-cost path in the CAG is going to represent an optimum degradation set. That is because it might be necessary to degrade multiple connections per link to find an optimum degradation set (and CAG cannot provide this solution). Thus, in the next section, we will provide an LP-based solution to be used when the CAG cannot find a \textit{good candidate set} and revenue functions are linear w.r.t. bandwidth.

\section{Provisioning under Resource Crunch} \label{sec:cag-and-alg}



\begin{figure} [t]
\centering
\resizebox{3.5in}{!}{
 \tikzstyle{branch}=[minimum size=3pt,inner sep=0pt]
 \begin {tikzpicture}
 
 \node (start)		[startstop] 							{Start};
 \node (crunched) 	[io, below = .4cm of start]	 			{Crunched Request $d_c$};
 \node (findCAG)		[process, below = .4cm of crunched]		{\specialcell[c]{I: CAG Provisioner\\(Algorithm \ref{alg:cheapest_cag_path})}};
 \node (CAGblock)	[decision, below = 0.4cm of findCAG]		{\specialcell[c]{Possible to\\serve $d_c$?}};
 \node (requestfits1)	[decision, below left = 0.4cm and 2cm of CAGblock]		{\specialcell[c]{$S^{good}$\\found?}};
 \node (findLP) 		[process, below = 0.8cm of CAGblock]	{\specialcell[c]{II: LP Provisioner}};
 \node (requestfits3) [decision, below = 0.4cm of findLP]		{\specialcell[c]{$S^{c}$\\found?}};
 \node (profitable1) [decision, below = 0.4cm of requestfits3]	{Profitable?};
 \node (serve) 		[process, below = 0.4cm of profitable1]	{\specialcell[c]{III: Throttle connections\\in degradation set\\ and allocate $d_c$}};
 \node (block) 		[process, right = 1cm of requestfits3]		{\specialcell[c]{Block $d_c$}};
 \node (invisible) 	[above = 1cm of profitable1]		{};
 
 \path [name path = a, arrow] (start) -- (crunched);
 \path [name path = a, arrow] (crunched) -- (findCAG);
 \path [name path = c, arrow] (findCAG) -- (CAGblock);
 \path [name path = d, arrow] (CAGblock.west) -| node [above, very near start] {\footnotesize yes} (requestfits1.north);
 \path [name path = e, arrow] (requestfits1) -| node [above, very near start] {\footnotesize no} (findLP);
 \path [name path = e, arrow] (requestfits1) |- node [left, very near start] {\footnotesize yes} (profitable1);
 \path [name path = g, arrow] (profitable1.south) -- node [left, very near start] {\footnotesize yes} (serve.north);
 \path [name path = h, arrow] (profitable1.east) -| node [above, very near start] {\footnotesize no} (block.south);
 \path [name path = i, arrow] (findLP) -- (requestfits3);
 \path [name path = j, arrow] (CAGblock.east) -| node [above, very near start] {\footnotesize no} (block.north);
 \path [name path = k, arrow] (requestfits3) -- node [above, very near start] {\footnotesize no} (block.west);
 \path [name path = l, arrow] (requestfits3.south) -- node [left, very near start] {\footnotesize yes} (profitable1.north);
  
 \end{tikzpicture}
}

\caption{PROVISIONER algorithm. All decision boxes are based on the outputs of the algorithms that precede them.}
\label{fig:algorithm}
\end{figure}
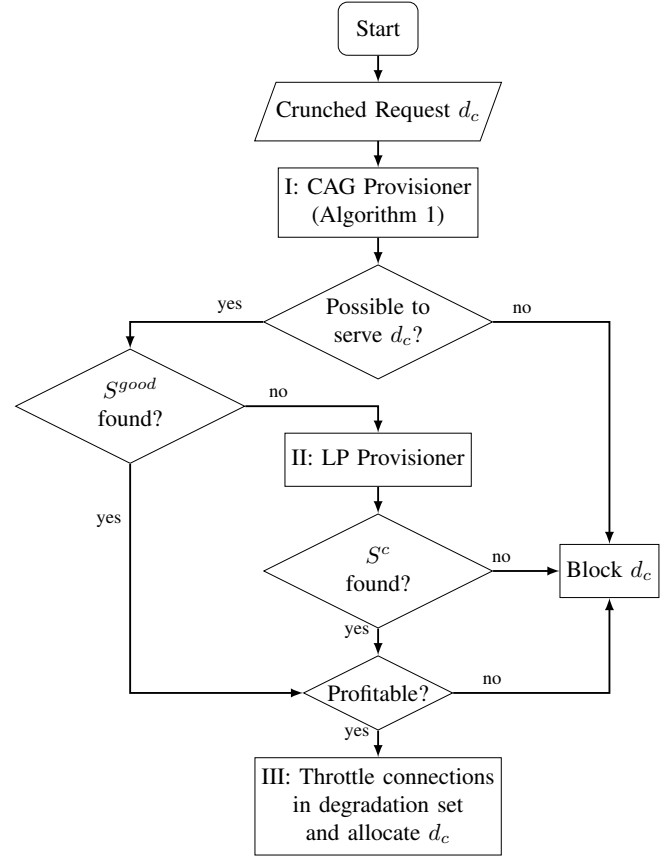

\subsection{PROVISIONER Algorithm}\label{sec:algorithm}

Fig. \ref{fig:algorithm} shows the high-level structure of the \textit{\underline{Provisio}ning U\underline{n}der R\underline{e}source C\underline{r}unch (PROVISIONER)} algorithm. Steps I, II, and III shown in the flowchart will be introduced in the following sections; however, we now give an overall explanation of how they fit together.

PROVISIONER starts with a crunched request $d_c$, and executes Algorithm I, which attempts to find a good candidate set of connections to degrade in order to serve request $d_c$. If this process does not find a good candidate set and revenue functions are linear w.r.t. bandwidth, the algorithm tries to find another candidate set through step II. If a candidate set is found in either of these steps, PROVISIONER checks if allocating $d_c$ by degrading the connections in the candidate set is profitable or not. If it is, these connections are degraded, and $d_c$ is allocated (III); if not, $d_c$ is blocked.

Aside from the flowchart of Fig. \ref{fig:algorithm}, the algorithm also uses a \textit{registry of degraded connections} (both served crunched requests and degraded connections) \footnote{In Section \ref{sec:results}, we also use the same registry for the other approaches in the comparisons.}. This registry is ordered first by the revenue of each connection (highest first); next, by the hop-length of each connection (fewer hops first). Thus, when an allocated connection departs, the degraded connections listed there are upgraded to a higher bandwidth, if possible, (i.e., to the highest possible bandwidth, up to the request's required bandwidth). With this registry, it is possible for a crunched request allocated at its minimum bandwidth to be subsequently upgraded to a higher bandwidth as soon as some connection departs.

\subsection{\textbf{I: CAG Provisioner}}\label{subsection:fitindegradable}

For each crunched request, Algorithm \ref{alg:cheapest_cag_path} tries to find min-cost paths in $CAG^{min}$ and $CAG^{req}$; decides whether to allocate $B_{c}^{min}$ or $B_{c}^{req}$ using Eqn. (\ref{equation_revloss}); checks if the candidate set found costs the same as the min-cost candidate set in the relaxed CAG returning $S^{good}$ if it does, or $S^c$ otherwise. 

\begin{algorithm}[!t]
\caption{CAG Provisioner}
\scriptsize
\label{alg:cheapest_cag_path}
 \begin{algorithmic}[1] 
 \renewcommand{\algorithmicrequire}{\textbf{Input:}}
 \renewcommand{\algorithmicensure}{\textbf{Output:}}
 \Require Crunched request $d_c$ and all CAGs
 \Ensure Block $d_c$, or Good set $S^{good}$, or Candidate set $S^{c}$, or null
 \State Find min-cost path $P^{*}_{min}$ in $CAG(N, s, t, B^{min}_{c})$ \label{alg:line:short_path}
 \If{$P^{*}_{min}$ not found}
 
 \Return Block $d_c$ \label{alg:line:short_path:not_found}
 \EndIf
 \State Find min-cost path $P^{*}_{req}$ in $CAG(N, s, t, B^{req}_{c})$ 
 \State Find min-cost path $P^{c}_{min}$ in $CAG^{min}(N, s, t, B^{min}_{c})$ 
 \State Find min-cost path $P^{c}_{req}$ in $CAG^{req}(N, s, t, B^{req}_{c})$
 \label{alg:line:short_capacitated_path}
 \If{$P^{c}_{min}$ and $P^{c}_{req}$ not found}
 
 	\Return null
 \EndIf
 
 \If{$\left( L(P^{c}_{req}) / B_{c}^{req} \right) \leq \left( L(P^{c}_{min}) / B_{c}^{min} \right)$}
 	\If{$L(P^{c}_{req}) = L(P^{*}_{req})$}
 
 	\Return $S^{good} =$ vertices of $P^{c}_{req}$
 \Else
 
 	\Return $S^{c} =$ vertices of $P^{c}_{req}$
 \EndIf
 \Else
 	\If{$L(P^{c}_{min}) = L(P^{*}_{min})$}
 
 	\Return $S^{good} =$ vertices of $P^{c}_{min}$
 \Else
 
 	\Return $S^{c} =$ vertices of $P^{c}_{min}$
 \EndIf
 \EndIf
 \end{algorithmic}
 \end{algorithm}
 
 \begin{table*}[t]\scriptsize
 \centering
 \caption{Linear Program Model Inputs.}
 \label{tab:inputs}
  \begin{tabularx}{\textwidth}{cX}
   \hline
   \hline
 	\textbf{Input}	&\textbf{Description} \\ \hline
 $d_c = <s, t, B_{c}^{min}, B_{c}^{req}, r_{c}, F_{c}>$ & Crunched request $d_c$: source $s$, destination $t$, minimum and required bandwidths $B_{c}^{min}$ and $B_{c}^{req}$, revenue function $r_{c}$.\\
 $P = \{(l_j, z_j)\}$ & List of links $l_j$ in the current path $P$. Each link $l_j$, has capacity $z_j$.\\
 $C = \{c_i = <B_{i}^{min}, B_{i}, r_{i}, V_{i}>\}$ & Set of connections that traverse at least some links of $P$. For each connection $c_i$: minimum required bandwidth $B_{i}^{min}$; currently allocated bandwidth $B_{i}$ (i.e., immediately before the 
crunched request arrived); revenue function $r_{i}$; and set $V_{i} = \{ v_{i}^{j} | v_{i}^{j} \in \{0,1\} \land l_j \in P\}$, where $v_{i}^{j}$ is $1$ if $c_i$ traverses link $l_j$ of $P$ (and 0, otherwise).\\
   \hline
  \end{tabularx}
 \end{table*}

As mentioned before, when free links exist, a dummy connection is added to the CAG. The cost of ``degrading'' dummy connections is zero, since they use free capacity. Our simulations show that it is common for a completely free path to exist between the crunched request's source and its destination (i.e., a path with some but not enough free capacity). This frequently leads to $L(P^{*}_{req})=0$ and $L(P^{*}_{min})=0$, which makes it rare for $S^{good}$ to be returned by Algorithm \ref{alg:cheapest_cag_path}. Also, if the $r_{c_{free}}$ are all zero, paths $P^{*}_{min}$, $P^{*}_{req}$, $P^{c}_{min}$, and $P^{c}_{req}$ tend to be long (which translates to long paths in the network). After testing several combinations, we concluded that setting $r_{c_{free}}$ to be the mean between the most expensive and the cheapest revenues $r_i$ leads to the best results (i.e., $S^{good}$ is frequently found and paths $P^{*}_{min}$, $P^{*}_{req}$, $P^{c}_{min}$, and $P^{c}_{req}$ are reasonably short). 

When Algorithm \ref{alg:cheapest_cag_path} finds a good degradation candidate set $S^{good}$, the operator can decide if serving the crunched request is a Profitable Decision. If $S^{c}$ is found, PROVISIONER continues with step II (described in Section \ref{subsection:notfitindegradable}), using $S^{c}$ as one of its inputs. It is also possible that no paths $P^{c}_{min}$ nor $P^{c}_{req}$ are found. In these cases, PROVISIONER will also try step II, however, without the input $S^{c}$. Finally, algorithm \ref{alg:cheapest_cag_path} might not find a path $P^{*}_{min}$ in the CAG (line \ref{alg:line:short_path:not_found}). If this happens, no possible degradation can be performed to allocate the crunched request, so it must be blocked.

The most demanding procedures of Algorithm \ref{alg:cheapest_cag_path} are four shortest-path computations on top of the CAG. In case Dijkstra is used, its worst-case is $O(|E| + V log V)$, where $V$ is the number of degradable connections in the network and $|E|$ is the number of edges in the CAG.

\subsection{\textbf{II: LP Provisioner}}\label{subsection:notfitindegradable}

If PROVISIONER reaches this point, Algorithm \ref{alg:cheapest_cag_path} was not able to find a good candidate set nor to assert that there is no possible degradation to execute (i.e., line \ref{alg:line:short_path:not_found} of Algorithm \ref{alg:cheapest_cag_path}). The optimum degradation set might either:
\begin{enumerate}[label=\textit{\alph*.}]
 \item Cost more than the min-cost path in the relaxed CAG; or
 \item Involve partially degrading different connections; or
 \item Not exist, because, even if all allocated connections in all links were throttled as much as possible (i.e., respecting the minimum required bandwidth of each connection), there would be no path with enough capacity.
\end{enumerate}

We propose an efficient LP-based algorithm to try to find a candidate degradation set in case Algorithm \ref{alg:cheapest_cag_path} did not find $S^{good}$ and revenue functions are linear w.r.t. bandwidth.



For this step, we only consider the physical network (the CAG is not used). We execute the LP model explained below for each of the k-shortest paths between the crunched demand's source and its destination (which may be precomputed), once for $B_{c}^{min}$ and once for $B_{c}^{req}$.
After considering all k paths, the algorithm chooses the cheapest path (and degradation set)  the LP model found (note that unlike the CAG, here we may partially degrade multiple connections to free up enough capacity on a link). Finally, it compares the cost of that set with the cost of the candidate set $S^c$ that Algorithm \ref{alg:cheapest_cag_path} (might have) found before. If the LP solution is cheaper, the algorithm calls it $S^c$ and returns it; otherwise, it returns the $S^c$ provided by Algorithm \ref{alg:cheapest_cag_path}. If no degradation is possible among the k-shortest paths and no $Sˆc$ was found before, the request is blocked.

The inputs of our LP model are described in Table \ref{tab:inputs}. For path $P$, with associated connections $C$, the LP uses a set of variables $\{y_{i}\}$, where $y_i$ represents the new bandwidth of each connection $c_i \in C$. We use $B_{c}$ to represent the allocated bandwidth to the new connection $d_c$ (i.e., depending on the execution, either $B_{c} = B_{c}^{min}$ or $B_{c} = B_{c}^{req}$). The model is as follows:
$$
\text{minimize } \left(X = \sum\limits_{{c_i} \in C} r_{i}(B_{i}) - \sum\limits_{{c_i} \in C} r_{i}(y_{i})\right)
$$
subject to
\begin{align}
B_{i}^{min} \leq y_{i} \leq B_{i}, && \forall {c_i} \in C \label{constraint:bandwidth}\\
\sum\limits_{{c_i} \in C} y_{i} \cdot v_{i}^{j} \leq z_j - B_{c}, && \forall l_j \in P\label{constraint:link-capacity}
\end{align}
Constraint \ref{constraint:bandwidth} enforces that connections' bandwidths will either be degraded (at most to the minimum required bandwidth of that connection), or will stay the same as before the crunched request arrived. Constraint \ref{constraint:link-capacity} enforces that each link in that path will have enough capacity to serve the crunched request's bandwidth $B_{c}$. Thus, this model might be infeasible if, even by degrading all connections in $C$, not enough capacity is freed for the crunched request $d_c$. If it is infeasible, the algorithm simply continues to the next path (or $d_c$ is blocked if it is infeasible for all k paths).

\subsection{\textbf{III: Throttling Connections}}

As Fig. \ref{fig:algorithm} illustrates, either if Algorithm \ref{alg:cheapest_cag_path} finds a good candidate set or if step II finds some candidate set, such set will be considered when deciding if serving the crunched request is profitable or not (as defined in Section \ref{ssec:definitions}).

Once it is determined that degrading the connections in the candidate set found is Profitable, the network operator can throttle the respective connections. The throttling operation is technology specific. For example, in case the network is OpenFlow-enabled version 1.2 and newer \cite{pfaff2012openflow}, this can be achieved by setting queues' maximum rates (e.g., field ``other-config:max-rate"). Other Layer-3 technologies also support equivalent rate-limiting \cite{Kumar:2015:BFH:2785956.2787478}. This throttling might also be achieved by reconfiguring lower layers, e.g., Bandwidth-Variable Transponders \cite{7045405} in Elastic Optical Networks. 

Once the connections in the candidate set are throttled, the operator can use a shortest-path algorithm (such as a capacitated version of Dijkstra) to find the shortest capacitated path that became available, and proceed to allocate the crunched request in such path. 

\section{CAG Space Complexity} \label{sec:complexity}

As the CAG represents the connections allocated in the network, its size can grow significantly. The number of vertices in the CAG is:
$$|V_{CAG}| = N_{c}^{d} + e_{free}$$ where $N_{c}^{d}$ is number of degradable connections in the network and $e_{free}$ is number of links with some free capacity.

Number of edges in the CAG depends on how many other connections share at least one physical node in their paths. If the network gets crowded by degradable connections, the CAG could become a dense graph in which case the number of edges is bounded by $|V_{CAG}|^2$.

In practice, the CAG size is rarely close to the aforementioned bounds. Usually, traffic that flows through core networks is combined into a smaller representation of the actual number of connections served --- which includes service aggregation. This results in CAGs that are easily manageable within memory. For a network managed/monitored by some system that records/manages the flows served (such as in an SDN environment), the number of vertices in the CAG can be the number of entries in the table that records such flows. Also, the computations performed on the CAG (aside from addition/removal of vertices) are shortest-path computations, which are fast even for large simple directed graphs.

As an example, during Resource Crunch, the average CAG has close to 270 vertices and $30,000$ edges, while 330 connections are served on average in our simulated Scenario C (which will be presented in the next section).



\section{Illustrative Numerical Examples} \label{sec:results}

We first introduce our pricing model; then, we discuss the simulation parameters and compared approaches; and, finally, we investigate results.

\subsection{Pricing Model}



We consider that the revenue generated by a connection depends on several factors:
\begin{enumerate}
\item bandwidth;
\item connection duration;
\item service class; and
\item square root of the length of the shortest path from the connection's source to its destination.
\end{enumerate}
As we will show, our solution tends to allocate requests through longer paths (when compared to other approaches). Thus, using a pricing model based on the shortest path (rather than the allocated path length) removes distortions that our model could have created otherwise. 

For purposes of our illustrative numerical examples, we consider the three following classes of service \cite{hong2013achieving}: 
\begin{enumerate}
\item \textit{\textbf{Interactive:}} These services directly impact end user experience (e.g., serving a user query), and they cannot suffer degradation. Also, they have the highest impact on revenue.

\item \textit{\textbf{Elastic:}} These services are more flexible than Interactive Services, and end users either have more flexibility in terms of their utilization experience (as when making a video call, or sending an e-mail), or are not directly impacted by them (as when replicating a data update between Data Centers). We assume these services can be degraded, and they have less impact on revenue than Interactive Services.

\item \textit{\textbf{Background:}} These services relate to maintenance activities that are not directly accessible to end users (backup migration, synchronization, configuration, etc). We consider that these services can be significantly degraded (more than Elastic Services), and they have the smallest impact on revenue.
\end{enumerate}

We also assume an SLA that imposes penalties if requests are blocked. We assume blocking costs amount to 50\% of the minimum revenue increase that would be generated from a request (i.e., considering how long the request would last at its minimum required bandwidth), except for Background traffic that has no blocking cost.

Thus, the revenue function of a connection $c_i$ is:
$$r_i(b_i, \theta_i, P_i^s, t_{init}, t_{end}) = b_i \cdot \theta_i \cdot \sqrt[]{len(P_i^s)} \cdot (t_{end} - t_{init})$$
where $b_i$ is bandwidth of $c_i$, $\theta_i$ is service class multiplier of that connection (shown in Table \ref{tab:traffic}), and $len(P_i^s)$ is the hop length of the shortest path $P_i^s$ between the connection's source and its destination.

The composition of the offered traffic in our simulation is shown in Table \ref{tab:traffic}. Every incoming request uses bandwidth that is uniformly distributed across the ranges presented.

\begin{table}[!h]
\caption{Service Classes and their Impacts on Revenue.}\label{tab:traffic}
\centering
\tabcolsep=0.11cm
 \begin{tabular}{|c|c|c|c|c|c|}
  \hline
	\specialcell[c]{Service\\Class}	&\specialcell[c]{\% of\\Total\\Traffic} &\specialcell[c]{Requested\\Bandwidth per\\Request (Gbps)}	&\specialcell[c]{Degradable\\bandwidth}	&\specialcell[c]{Service Class\\Multiplier $\theta$\\$\left(\sfrac{\mbox{\$}}{(\mbox{Gbit})}\right)$}	&\specialcell[c]{Cost of\\Blocking\\$\left(\sfrac{\mbox{\$}}{(\mbox{Gbit})}\right)$}\\ \hline
	Inter.	&20\%	&$U(0.1, 4.0)$	& 0.00\%	&0.08	&0.04\\ \hline
	Elastic		&30\%	&$U(0.1, 6.0)$	& 33.3\%	&0.06	&0.03\\ \hline
	Backgr.	&50\%	&$U(0.1, 8.0)$	& 50.0\%	&0.04	&0.00\\ \hline
  \hline
 \end{tabular}
 
\end{table}


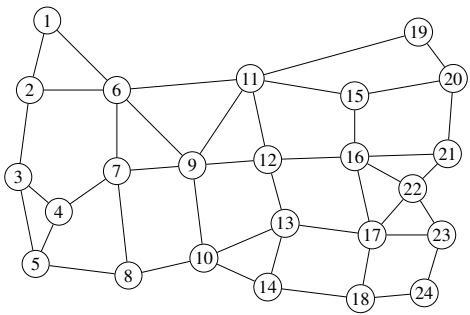
\begin{figure}[!t]
 \centering
 \resizebox{2.5in}{!}{
 \begin{tikzpicture}
 
 \node[draw,font=\small,circle,fill=white,minimum size=1em,inner sep=1] (12) at (0,0) {12};
 \node[draw,font=\small,circle,fill=white,minimum size=1em,inner sep=1] (16) at (1.5,.05) {16};
 \node[draw,font=\small,circle,fill=white,minimum size=1em,inner sep=1] (15) at (1.5,1.1) {15};
 \node[draw,font=\small,circle,fill=white,minimum size=1em,inner sep=1] (21) at (3.1,.1) {21};
 \node[draw,font=\small,circle,fill=white,minimum size=1em,inner sep=1] (11) at (-.3,1.4) {11};
 \node[draw,font=\small,circle,fill=white,minimum size=1em,inner sep=2] (9) at (-1.3,-.1) {9};
 \node[draw,font=\small,circle,fill=white,minimum size=1em,inner sep=2] (7) at (-2.6,-.2) {7};
 \node[draw,font=\small,circle,fill=white,minimum size=1em,inner sep=2] (6) at (-2.6,1.2) {6};
 \node[draw,font=\small,circle,fill=white,minimum size=1em,inner sep=2] (8) at (-2.4,-2) {8};
 \node[draw,font=\small,circle,fill=white,minimum size=1em,inner sep=2] (5) at (-4,-1.8) {5};
 \node[draw,font=\small,circle,fill=white,minimum size=1em,inner sep=2] (3) at (-4.3,-.3) {3};
 \node[draw,font=\small,circle,fill=white,minimum size=1em,inner sep=2] (2) at (-4.1,1.2) {2};
 \node[draw,font=\small,circle,fill=white,minimum size=1em,inner sep=2] (1) at (-3.8,2.4) {1};
 \node[draw,font=\small,circle,fill=white,minimum size=1em,inner sep=2] (4) at (-3.6,-.9) {4};
 \node[draw,font=\small,circle,fill=white,minimum size=1em,inner sep=1] (10) at (-1.1,-1.7) {10};
 \node[draw,font=\small,circle,fill=white,minimum size=1em,inner sep=1] (13) at (.3,-1.1) {13};
 \node[draw,font=\small,circle,fill=white,minimum size=1em,inner sep=1] (17) at (1.8,-1.3) {17};
 \node[draw,font=\small,circle,fill=white,minimum size=1em,inner sep=1] (23) at (3,-1.3) {23};
 \node[draw,font=\small,circle,fill=white,minimum size=1em,inner sep=1] (22) at (2.5,-.5) {22};
 \node[draw,font=\small,circle,fill=white,minimum size=1em,inner sep=1] (20) at (3.2,1.4) {20};
 \node[draw,font=\small,circle,fill=white,minimum size=1em,inner sep=1] (19) at (2.6,2.2) {19};
 \node[draw,font=\small,circle,fill=white,minimum size=1em,inner sep=1] (24) at (2.7,-2.3) {24};
 \node[draw,font=\small,circle,fill=white,minimum size=1em,inner sep=1] (18) at (1.6,-2.4) {18};
 \node[draw,font=\small,circle,fill=white,minimum size=1em,inner sep=1] (14) at (0,-2.2) {14};
  
 \draw (6) -- (9);
 \draw (12) -- (16); 
 \draw (12) -- (13);
 \draw (12) -- (9);
 \draw (21) -- (16);
 \draw (13) -- (17);
 \draw (16) -- (17);
 \draw (23) -- (22);
 \draw (16) -- (22);
 \draw (17) -- (22);
 \draw (22) -- (21);
 \draw (8) -- (10);
 \draw (10) -- (14);
 \draw (10) -- (13);
 \draw (13) -- (14);
 \draw (7) -- (8);
 \draw (7) -- (9);
 \draw (18) -- (14);
 \draw (10) -- (9);
 \draw (17) -- (18);
 \draw (23) -- (24);
 \draw (24) -- (18); 
 \draw (17) -- (23);
 \draw (5) -- (8);
 \draw (4) -- (5);
 \draw (3) -- (4);
 \draw (3) -- (2);
 \draw (2) -- (6);
 \draw (2) -- (1);
 \draw (6) -- (7);
 \draw (1) -- (6);
 \draw (6) -- (11);
 \draw (9) -- (11);
 \draw (4) -- (7);
 \draw (11) -- (15);
 \draw (15) -- (16);
 \draw (15) -- (20);
 \draw (11) -- (19);
 \draw (11) -- (12);
 \draw (19) -- (20);
 \draw (20) -- (21);
 \draw (3) -- (5);
 
  \end{tikzpicture}
  }
 \caption{24-node US-wide topology used in this study.}
 \label{fig:USNet}
 
\end{figure}

\begin{figure}[!t] 
\centering
 \subfloat[\label{fig:rc_profile:a}] {
 \includegraphics[height=1.5in]{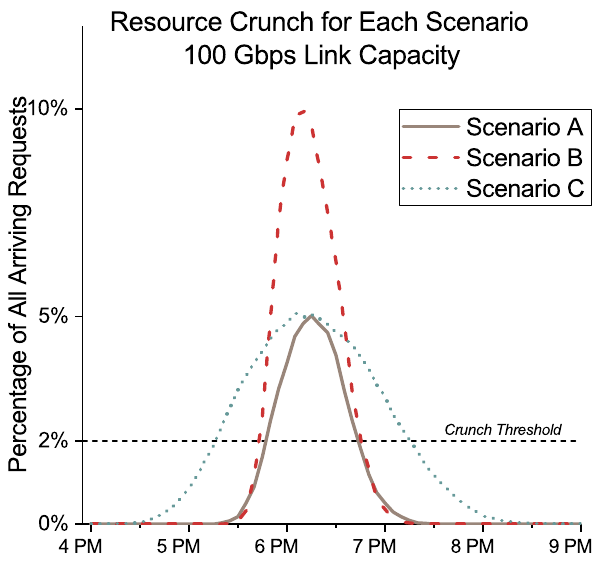} 
 } \subfloat[\label{fig:rc_profile:b}] {
 \includegraphics[height=1.5in]{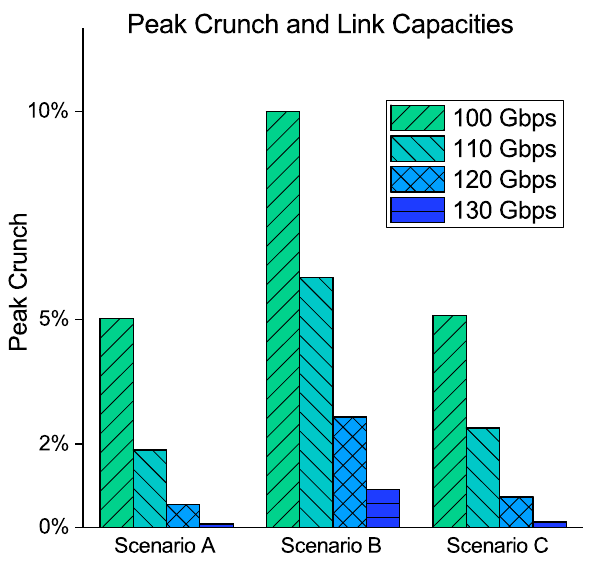} 
 }
 \caption{Percentage of crunched requests for each scenario under the \textbf{Baseline} approach. In (a), daily Resource Crunches for each scenario. In (b), how the peak Resource Crunch changes for higher network capacities.}
 \label{fig:rc_profile} 
\end{figure}

\subsection{Simulation Settings} \label{sec:simset}

A dynamic network simulation was implemented to evaluate the performance of the proposed algorithm. We analyzed Resource Crunch caused by traffic fluctuations as follows:
\begin{enumerate}[label=\textit{\alph*.}]
 \item New requests arrive with exponential inter-arrival times. Mean inter-arrival times vary throughout the day to reflect daily traffic fluctuations, in a cyclical pattern similar to that of Fig. 1. For simplicity, we consider that these variations have a sinusoidal shape;
 \item All connections have exponentially-distributed durations of mean 30 minutes;
 \item Requests uniformly select a source-destination pair and ask for a certain amount of bandwidth (see Table \ref{tab:traffic}).
\end{enumerate}

The network was considered to be under Resource Crunch when more than two percent of requests were crunched. For each result, $10,000$ days were simulated, and Resource Crunch occurred every day. The 95\% Confidence Interval of the results presented is of $\pm 0.05 \%$ of their means (prior to the normalization in case of Figs. \ref{res:revenue_profit} and \ref{res:percent_crunch}). These intervals are not shown because they are too small to be displayed.

We used the topology of Fig. \ref{fig:USNet} and the following scenarios:
\begin{itemize}
 \item \textbf{Scenario A}: daily Resource Crunch that lasts a little over one hour, and, at its peak, causes five percent of requests to be crunched;
 \item \textbf{Scenario B}: daily Resource Crunch that lasts a little over one hour, and, at its peak, causes ten percent of requests to be crunched; and
 \item \textbf{Scenario C}: daily Resource Crunch that lasts a little over two hours, and, at its peak, causes five percent of requests to be crunched.
\end{itemize}

Fig. \ref{fig:rc_profile:a} shows the ratio of crunched requests for the topology of Fig. \ref{fig:USNet} with links at 100 Gbps. Fig. \ref{fig:rc_profile:b} shows how the peak Resource Crunch would change if links' capacities were higher. At 130 Gbps, Resource Crunch does not occur. 

Under each scenario, these approaches were compared:
\begin{enumerate}
\item \textbf{100 Gbps Baseline}: in this approach, if a request is crunched, it is blocked. This is the approach used to generate the curves of Fig. \ref{fig:rc_profile}.
\item \textbf{130 Gbps Baseline}: in this approach, if a request is crunched, it is blocked. This is the only approach that uses a different link capacity, namely, 130 Gbps. As Fig. \ref{fig:rc_profile:b} shows, this approach does not experience Resource Crunch. In other words, this approach allows us to understand how much revenue could be generated if the network was able to absorb all the traffic that is being offered. Note, however, that this hypothetical 30\%-higher-capacity network would require significant Capital Expenditures, which would likely not be justified by the revenue increase it enables.
\item \textbf{PROVISIONER-k1}: approach proposed in our study. If the crunched request reaches step II (Fig. \ref{fig:algorithm}), $k=1$ path is investigated (i.e., only the shortest path).
\item \textbf{PROVISIONER-k10}: approach proposed in our study. If the crunched request reaches step II (Fig. \ref{fig:algorithm}), $k=10$ paths are investigated.
\item \textbf{LP-k1}: this approach consists of only executing step II (Fig. \ref{fig:algorithm}) when a request is crunched (without going through the entire flow of PROVISIONER, i.e., without executing Algorithm \ref{alg:cheapest_cag_path}). Only shortest path is investigated to find a cheap degradation set to accommodate the crunched request.
\item \textbf{LP-k10}: this approach consists of only executing step II (Fig. \ref{fig:algorithm}) when a request is crunched. $k=10$ shortest paths are investigated.
\item \textbf{LP-k100}: this approach consists of only executing step II (Fig. \ref{fig:algorithm}) when a request is crunched. $k=100$ shortest paths are investigated. This approach tends to find optimum degradation sets, since it investigates a large number of paths.
\item \textbf{SP-k10}: a similar approach as the one described in \ref{ssec:sp_approach} (which is also similar to \cite{savas2016backup, 7842043, roy2014network}). Instead of only considering the single shortest path, this approach goes from the shortest to the $k^{th}$ shortest path and, for each path, degrades the cheapest connections that lie in it until enough capacity is freed for the crunched request. As soon as a path frees up enough capacity for the request, it checks if such degradations are Profitable. If so, the crunched request is served (by throttling the other connections); otherwise, it is blocked.
\end{enumerate}
Also, the registry of degraded requests explained at the end of Section \ref{ssec:problem-statement} is utilized by all approaches.

\subsection{Profit and Revenue Impacts}

Table \ref{tab:rev_rc} shows the average profits generated during Resource Crunch for each approach in each scenario. Scenario A goes through a less intense Resource Crunch than Scenario B, and both of them go through shorter Resource Crunch than Scenario C. The revenues generated in each of them vary accordingly. PROVISIONER performs better than the other options in all situations. Differences between the approaches and scenarios will be further analyzed in this section.

\begin{table}[!h]
\caption{Average Profits During Resource Crunch.}\label{tab:rev_rc}
\centering
\tabcolsep=0.11cm
 \begin{tabular}{|c|c|c|c|}
  \hline
	Approach		& Scenario A (\$) & Scenario B (\$) & Scenario C (\$) \\ \hline
 130 Gbps Baseline	&253271.14 	& 323723.38 	& 586828.17	\\ \hline
	100 Gbps Baseline	&240168.01 	& 292761.09 	& 548615.07	\\ \hline
	PROVISIONER-k1		&246347.08 	& 308631.33 	& 567065.61	\\ \hline
	PROVISIONER-k10		&246573.22 	& 308564.16 	& 566988.05	\\ \hline
 LP-k1				&245630.19 	& 307645.79 	& 565478.59	\\ \hline
 LP-k10				&245235.01 	& 307964.51 	& 565552.51	\\ \hline
 LP-k100				&245886.76 	& 292382.51		& 564464.21	\\ \hline
 SP-k10				&240640.74 	& 307329.96		& 548971.86	\\ \hline
  \hline
 \end{tabular}
 
\end{table}

The results of Table \ref{tab:rev_rc} refer to the duration of the Resource Crunch experienced by the 100 Gbps Baseline approach. As we will further investigate, different approaches lead the network to higher utilization levels, which end up elongating the Resource Crunch duration. Not only that, but also requests allocated during Resource Crunch may still be present after the Crunch is over, potentially affecting the performance of the network throughout the day. Thus, in the next results, we focus on the effects of each approach throughout the day.

\begin{figure}[!t]
\centering
 \subfloat[\label{sres:revenue}] {
 \includegraphics[width=0.8\linewidth]{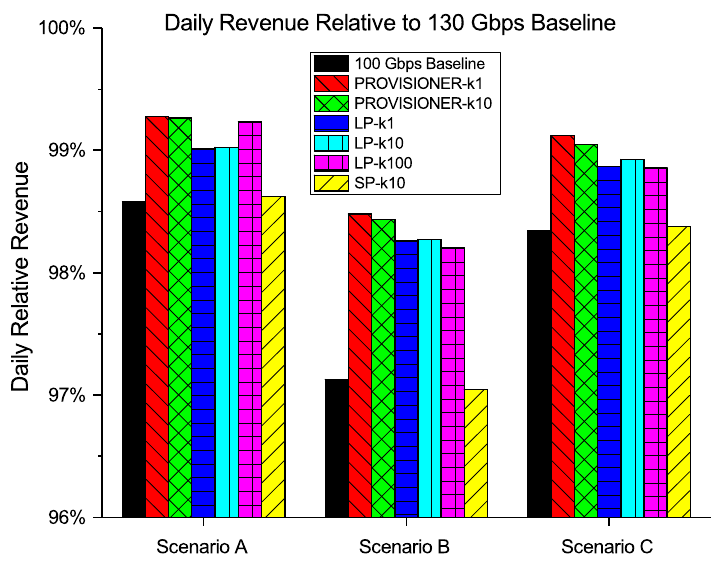} 
 }
 \\
 \hspace{-0.5cm}
 \centering
 \subfloat[\label{sres:profit}] {
 \includegraphics[width=0.8\linewidth]{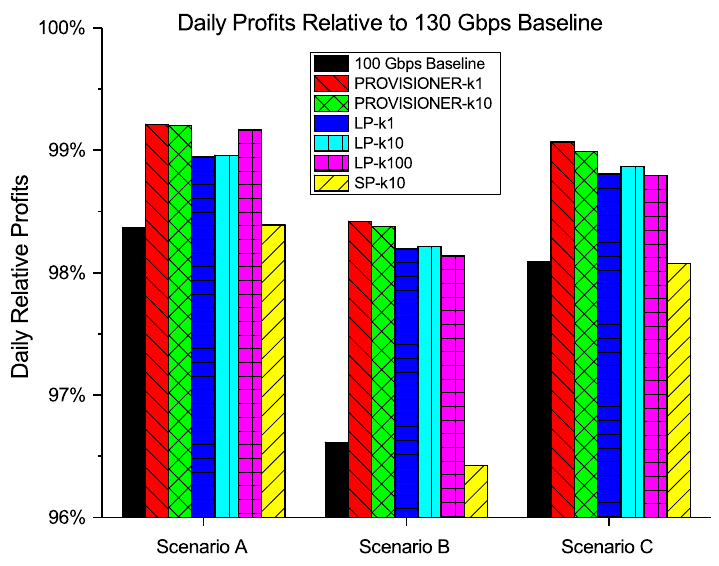} 
 }
 \caption{In (a), how PROVISIONER contributes to daily revenue. In (b), how PROVISIONER contributes to daily profits (i.e., revenue minus cost of blocking requests).}
 \label{res:revenue_profit} 
\end{figure}

In a daily perspective, Fig. \ref{res:revenue_profit} shows how each approach performs relative to the 130 Gbps Baseline; in Fig. \ref{sres:revenue}, with regards to revenue increases, and, in Fig. \ref{sres:profit}, with regards to profit increases (measured as revenue minus cost of blocking requests). PROVISIONER performs better than the other approaches in two ways: it generates more revenue; and it incurs lower blocking costs, resulting in higher profits. This behavior, however, is related to the duration and intensity of the Resource Crunch (i.e., to each scenario). 



In Scenario A, the network goes through a shorter and less intense Crunch, thus, network resources do not get extremely congested. This is also evidenced by the 100 Gbps Baseline performing better in this scenario than it does in the other scenarios. As a result, it is more common that crunched requests can be served by degrading up to one connection per link. This allows PROVISIONER to find good solutions frequently. Accordingly, in such scenario, Fig. \ref{res:revenue_profit} shows that LP-k100 matches the behavior of PROVISIONER and both perform much better than the other LP-based solutions as well as SP-k10.

In Scenario B, the duration of Resource Crunch is similar to Scenario A, however, it sees a much higher ratio of crunched requests. On the one hand, this incurs in a higher number of crunched requests (i.e., PROVISIONER is executed more often than in Scenario A). This is reflected in that the performance of PROVISIONER and the LP-based approaches is much higher than the 100 Gbps Baseline, when compared to the other scenarios. On the other hand, the network becomes more occupied than in Scenario A. As a result, under such high occupation, shorter paths are more beneficial (since long paths result in higher network utilization and, hence, more crunched requests). This is reflected not only in the fact that PROVISIONER-k1 performs better than PROVISIONER-k10 (in Scenario B), but also that LP-k100 performs worse than the other LP-based approaches (i.e., it over-optimizes the results for individual requests, and not for the overall Resource Crunch period). Nevertheless, on occasion, it is still beneficial to find longer paths to serve some requests. This is why PROVISIONER performs better than the other approaches.

In Scenario C, the duration of Resource Crunch is longer than in Scenario A, but has a similar peak of crunched requests. Because this scenario lasts longer, it has more requests, which allows for higher revenues and profits than Scenario A, as seen in the difference between the 100 Gbps Baseline approach and the others in Scenario C of Fig. \ref{res:revenue_profit}. However, this scenario does not generate such high gains (when compared to the 100-Gbps Baseline) as the approaches in Scenario B.
This is because, during Resource Crunch, all approaches increase network utilization when compared to the 100 Gbps Baseline, since crunched requests are served through potentially long paths and these requests have non-negligible holding times. This leads to more crunched requests in the future, which leads to even higher utilization levels, in cyclical manner. If the offered load decreases soon (as in the other scenarios), this cycle is broken, and the network goes back to normal operation\footnote{Note that a crunched request is bound to generate less profits than a similar request that can be served normally.}. Thus, a long Resource Crunch as that of Scenario C tends to occupy the network a lot, which translates to shorter paths being beneficial. Thus PROVISIONER-k1 performs considerably better than PROVISIONER-k10. Also, the performances of the other approaches get closer. As before, though, it is still beneficial to find optimum solutions, which is why PROVISIONER performs better than the others.

\begin{figure}[!t] 
 \centering
 \includegraphics[width=0.8\linewidth]{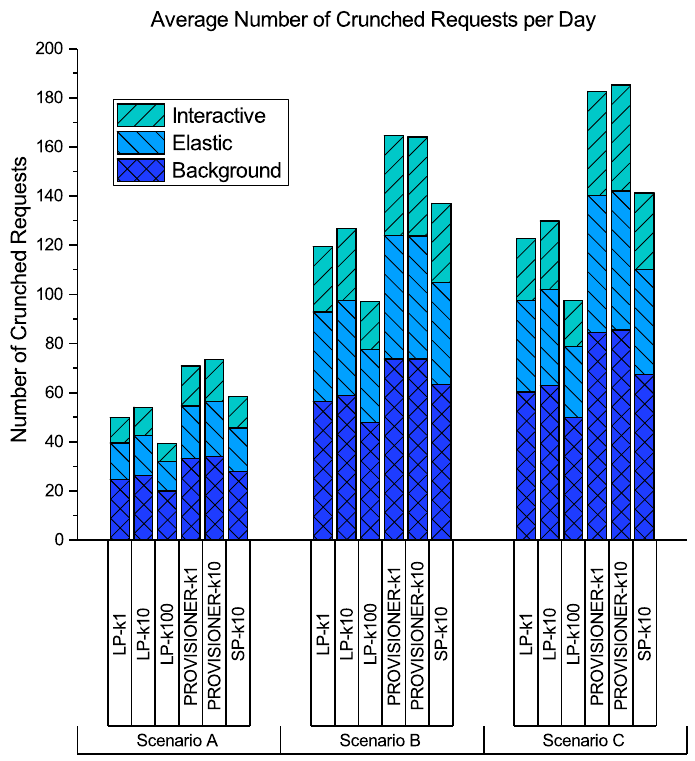} 
 \caption{Number of crunched requests and their service classes in each scenario.}
 \label{res:num_crunch} 
\end{figure}

\begin{figure}[!t] 
 \centering
 \includegraphics[width=1.0\linewidth]{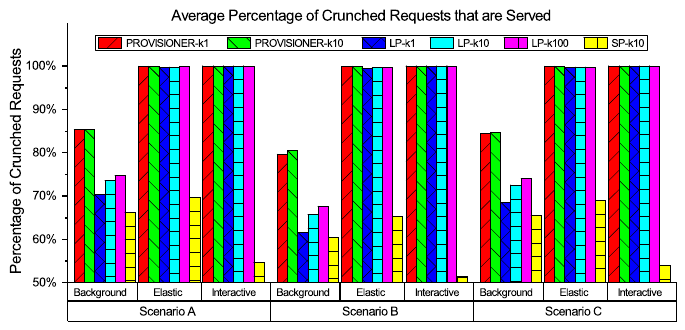} 
 \caption{Acceptance ratio of crunched requests in each scenario.}
 \label{res:percent_crunch} 
\end{figure}

In all scenarios, SP-k10 has a poor overall performance. Since it tries to degrade the cheapest connections, it tends to degrade shorter ones. With that, each crunched request only frees enough capacity to place itself. After a while, the network gets crowded with crunched requests and loses overall degradable capacity. Thus, new crunched requests are blocked due to a lack of degradable capacity.

\begin{figure}[!t] 
 \centering
 \includegraphics[width=0.8\linewidth]{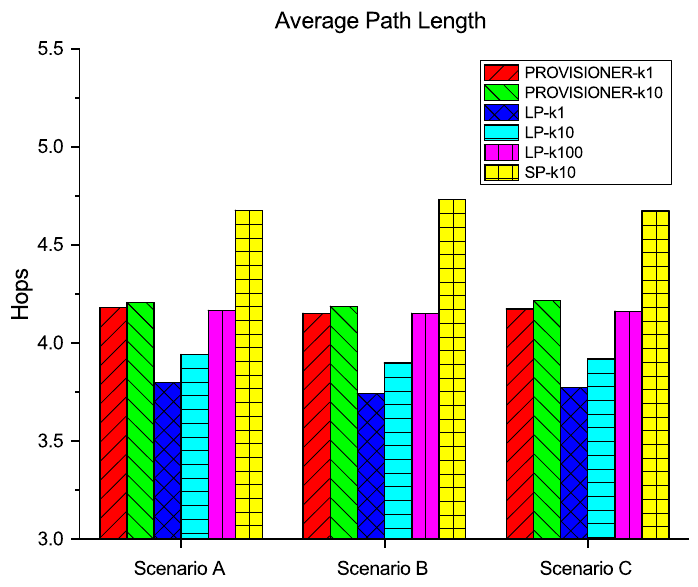} 
 \caption{Average path length.}
 \label{res:path} 
\end{figure}

Fig. \ref{res:num_crunch} shows the number of requests that are crunched on an average day for each scenario. Note that Interactive requests tend to require less bandwidth, followed by Elastic, followed by Background (see Table \ref{tab:traffic}). This is directly related to the number of requests from each service class that are crunched, as shown in Fig. \ref{res:num_crunch}. As stated before, the lengthy Resource Crunch of Scenario C and the intense Resource Crunch of Scenario B tend to cause more requests to be crunched than in Scenario A. This is particularly true under the PROVISIONER approach, mostly because it achieves a higher network utilization (as will be explored in Fig. \ref{res:path}). Note, however, that more crunched requests are not the reason for the better performance of PROVISIONER, because, if the network utilization level was lower, fewer requests would be crunched, and it is better to have fewer crunched requests.

Fig. \ref{res:percent_crunch} shows the average percentage of crunched requests served in each scenario by each approach. Interactive and Elastic requests (i.e., the most expensive ones) tend to always be served either by PROVISIONER or by the LP-based solutions. However, PROVISIONER is able to serve a much higher percentage of Background traffic. That is because PROVISIONER is able to find extremely cheap degradation sets (through the CAG). This is particularly important for Background traffic, because these requests offer very little revenue increases. PROVISIONER is, thus, fairer to these lower-cost requests without hurting other high-cost ones (as it allocates close to a 100\% of Elastic and Interactive requests). 


\begin{table}[!t]
\caption{Average Execution Time.}\label{tab:exec_time}
\centering
\tabcolsep=0.11cm
 \begin{tabular}{|c|c|}
  \hline
	Approach		& Execution Time (ms) \\ \hline
	PROVISIONER-k1	&7.30	\\ \hline
	PROVISIONER-k10	&31.54	\\ \hline
 LP-k1			&7.95	\\ \hline
 LP-k10			&43.68	\\ \hline
 LP-k100			&469.90	\\ \hline
 SP-k10			&7.32	\\ \hline
  \hline
 \end{tabular}
 
\end{table}

Fig. \ref{res:path} shows the average path length of the crunched requests served by each approach. There is no significant difference among the scenarios. However, note that PROVISIONER tends to serve requests through paths longer than those of LP-k1 and LP-k10. The LP-k100 solution tends to find paths of similar length, as it tends to find optimum solutions. The SP-k10 solution, due to its greediness, tends to find longer paths, because those are the ones it can find that can both free enough capacity for the crunched request and are cheap enough for the crunched request to be served. 

The longer paths found by PROVISIONER along with the higher acceptance ratio of crunched requests, tends to elevate the network utilization. This contributes to the elongation of Resource Crunch. However, as can be seen in the various results, PROVISIONER can deal with that very well, generating higher profits than all the other approaches. Not only that, but PROVISIONER is also faster than the others. Consider Table \ref{tab:exec_time} (run on a Core i7 16GB RAM machine): PROVISIONER-k1 runs even faster than LP-k1. This is because, many times, PROVISIONER does not have to solve the LP model (and Algorithm \ref{alg:cheapest_cag_path} is much faster). Thus, PROVISIONER can outperform the other approaches at much faster execution times, while, also, being fairer to low cost requests than the others.

\section{Conclusion} \label{sec:conclude}

In this work, we observed that increasing average network utilization might lead to the occurrence of Resource Crunch. In such situation, it is important to have an efficient method to decide how to throttle connections to serve incoming requests that would otherwise be blocked. We showed how the use of bandwidth-flexible requests helps in dealing with Resource Crunch. We introduced the Connection Adjacency Graph (CAG), a useful tool to represent the network state. We showed how the decision of whether or not to serve a crunched request (and, if so, where to allocate it) is challenging. We developed the PROVISIONER algorithm which is based on two methods: one using the CAG; and an efficient Linear Program (LP). We compared the results of our method with LP-based approaches and an existing greedy approach and showed how our method outperforms them. The results confirmed that PROVISIONER is efficient in maximizing profits. They also showed that it provides high acceptance rates for low-paying requests without detriment to high-paying requests. Finally, we showed how PROVISIONER executes faster than the other approaches.

Future work includes analyzing at what point network upgrade should be performed, and, in a scenario with requests that have malleable start/end times, how to schedule such requests during Resource Crunch.

\section{Acknowledgments}\label{sec:Ack}

We thank the anonymous reviewers for their constructive comments in improving the paper. R. Louren\c{c}o was funded by CAPES Foundation (Proc. 13220-13-6). M. Tornatore acknowledges the research support from COST Action CA15127. This work was supported in part by DTRA grant HDTRA1-14-1-0047. 

\bibliographystyle{IEEEtran}
\bibliography{IEEEabrv,ResourceCrunch.bib}

\end{document}